\documentclass[doublespace]{qjrms4}

\usepackage{tikz}
\usepackage{natbib}
\newdimen\alignnumbersep
\usepackage{etoolbox}
\makeatletter
\patchcmd{\math@cr@@@align@measure}
     {\df@tag\fi\fi}
     {\df@tag\fi\fi\kern\alignnumbersep}
     {}{}
\patchcmd{\math@cr@@@align}
    {\make@display@tag}
    {\make@display@tag\kern\alignnumbersep}
    {}{}
\makeatother

\newcommand{\ISS}{ISS$_{rev}$ }
\newcommand{\ISSq}{ISS$_{quant}$ }

\begin{document}
\runningheads{M. Weniger, F. Kapp and P. Friederichs}{Spatial Verification Using Wavelet Transforms: A Review}
\title{Spatial Verification Using Wavelet Transforms: A Review}
\author{Michael Weniger\affil{a}\corrauth,
Florian Kapp\affil{a} and Petra Friederichs\affil{a}}
\address{\affilnum{a}Meteorological Institute, University of Bonn, Germany.}
\corraddr{Michael Weniger, Meteorological Institute, University of Bonn, 
				Auf dem H\"ugel 20, 53121 Bonn, Germany. E-mail: mweniger@uni-bonn.de}

\begin{abstract}
Due to the emergence of new high resolution numerical weather prediction (NWP) models and the availability of new or more reliable remote sensing data, the importance of efficient spatial verification techniques is growing. Wavelet transforms offer an effective framework to decompose spatial data into separate (and possibly orthogonal) scales and directions. Most wavelet based spatial verification techniques have been developed or refined in the last decade and concentrate on assessing forecast performance (i.e. forecast skill or forecast error) on distinct physical scales. Particularly during the last five years, a significant growth in meteorological applications could be observed. However, a comparison with other scientific fields such as feature detection, image fusion, texture analysis, or facial and biometric recognition, shows that there is still a considerable, currently unused potential to derive useful diagnostic information. In order to tab the full potential of wavelet analysis, we revise the state-of-the art in one- and two-dimensional wavelet analysis and its application with emphasis on spatial verification. We further use a technique developed for texture analysis in the context of high-resolution quantitative precipitation forecasts, which is able to assess structural characteristics of the precipitation fields and allows efficient clustering of ensemble data.
\end{abstract}
\keywords{spatial verification; forecast skill; forecast error; scale separation; wavelet; DWT; CWT; review}

\maketitle

\clearpage
\section{Introduction}
\label{sec:introduction}
The emergence of high-resolution numerical weather prediction (NWP) models presents new challenges for forecast verification. Traditional pointwise verification methods experience several complications and might yield misleading results \citep{gilleland2013testing}. First, small-scale fluctuations tend to dominate these scores, thereby missing important information on medium and large scales. Second, the displacement of features is penalized twice, once for missing the observed feature at a certain gridpoint, and once for wrongly forecasting a feature at a second gridpoint. This effect is known as \textit{double penalty}. These issues may cause high-resolution models to exhibit worse verification scores than lower-resolution models, even in cases where they are more realistic and useful. A third shortcoming of pointwise scores accumulated over a whole field concerns the calculation of confidence intervals, which typically assumes that the field values are statistically independent and identically distributed. Unfortunately, this does not hold true for most meteorological variables, particularly in a high-resolution environment, and leads to an overestimation of sample sizes and consequently to confidence intervals that are too narrow. Furthermore, traditional scores suffer from a lack of important diagnostic information about the type of errors, e.g.\ displacement errors, wrong structure / texture of features or skillful spatial scales.

These issues as well as the availability of new or more reliable remote sensing data have lead to the development of new verification methods for spatial fields that are able to evaluate output of high-resolution models based on observed spatial fields. Spatial verification methods can roughly be divided into four categories: neighborhood methods or \textit{fuzzy} verification \citep{Theis05,roberts2008scale,mittermaier2013long,skok2015analysis}, scale separation \citep{briggs1997wavelets,harris2001multiscale,casati2004wavelet,lack2010object}, field deformation \citep{keil2007displacement,keil2009das,gilleland2010image} and feature based methods \citep{ebert2000verification,davis2006object,wernli2008sal, weniger2015sal}. {While different techniques concentrate on different diagnostic information, they all share a common denominator in recognizing the spatial correlations in the data. Hence, they are able to provide more reasonable confidence intervals than pointwise verification methods \citep{davis2009method,hering2011comparing,gilleland2013testing}. We refer to \cite{ebert2008}, \cite{gilleland2009, Gilleland2010} and \cite{ebert2013progress} for a thorough overview of spatial verification methods. 

The present study takes a closer look at scale separation techniques based on wavelet transforms, which are a popular tool in other scientific disciplines such as image processing and offer a number of advantages to assess the forecast skill or error for spatial fields. First, wavelets have proven to be very efficient in data reduction, which is crucial for large sets of high-resolution meteorological data. Second, they offer a framework for an orthogonal scale decomposition, which naturally leads to the decomposition of traditional pointwise verification measures such as the mean square error (MSE). This allows, for instance, the evaluation of a model on isolated physical scales. Third, wavelets are localized in both time and frequency and therefore, contrary to Fourier based approaches, do not require stationarity of the data. Fourth, wavelet transforms work well in noisy environments, which is particularly important for remotely sensed data with non-negligible observational uncertainties. Furthermore, wavelets bear the potential to bridge the gap between neighborhood and feature based methods: in terms of filtering the low-pass filter of a wavelet transform corresponds to (directionally weighted) box averaging, whereas the results of the high-pass filter can be used for feature extraction and key point detection \citep{fauqueur2006multiscale}.

While promising wavelet-based methods for (high-resolution) spatial verification have been developed in the last decade, a comparison to related scientific fields such as spatial bootstrapping to generate reliable confidence intervals \citep{solow1985bootstrapping,breakspear2004spatiotemporal,whitcher2006wavelet,csendur2007resampling}, denoising \citep{mihcak1999low,chang2000adaptive,buades2005review,pizurica2006review,chen2013wavelet}, feature detection \citep{mallat1992characterization,wang2012review,yan2014wavelets,pimentel2014review}, image fusion \citep{li1995multisensor,pauly2009wavelet,petrosian2013wavelets,suraj2014discrete}, texture analysis \citep{chang1993texture,unser1995texture,prats2011multivariate,hsin2012new,virmani2013svm}, facial and biometric recognition \citep{daugman1993high,boles1998human,liu2002gabor,daugman2007,cao2012illumination} or data compression \citep{coifman1992entropy,christopoulos2000jpeg2000,li2011remote,bayazit2011adaptive,taubman2012jpeg2000} reveals an immense and currently unused potential.

In order to tab the full potential of wavelet analysis, we revise the state-of-the art in one- and two-dimensional wavelet analysis and its application with emphasis on spatial verification. We further use a technique developed for texture analysis in the context of high-resolution quantitative precipitation forecasting.

This paper is thus structured as follows.
The mathematical framework for one- and two-dimensional continuous and discrete wavelet transforms is provided in section~\ref{sec:mathematical-framework}. A comprehensive review of existing spatial verification techniques based on wavelet decomposition is given in section~\ref{sec:spatial-verification-techniques} followed by an overview of meteorological applications in section~\ref{sec:applications}. An exemplary application of texture analysis \citep{eckley2010locally} to meteorological data is presented in section~\ref{sec:texture}.

\section{Mathematical Framework}
\label{sec:mathematical-framework}

\subsection{One-Dimensional Wavelet Transforms}
\label{subsec:1d-wavelet-transform}
Let us first consider the one-dimensional, real-valued case. Mathematical literature such as \cite{daubechies1992ten} assume that the signal to be analyzed is given in the form of an integrable function $f(t)$. This abstract formulation includes discrete one-dimensional sets of data via stepwise constant functions. Since the data in (spatial) verification are usually discrete with a finite resolution, we follow \cite{torrence1998practical} and denote the data, e.g. a time series, by $X=\{x_k\}_{k=1}^N\in\mathbb{R^N}$ with $N\in\mathbb{N}$. Assume that the time series has equal time-spacing $\Delta T>0$. A wavelet is a function $\psi:\mathbb{R}\rightarrow\mathbb{R}$, which is localized in time and frequency (for mathematical details we refer to \cite{daubechies1992ten} or \cite{farge1992wavelet}) and has vanishing mean, i.e.
\begin{align}
	\int_{\mathbb{R}} \psi(t) dt = 0.
\end{align}
Such a function is called a \textit{mother wavelet}, because it is the origin for a family of scaled and shifted wavelets
\begin{align}
	\psi_{s,l}(t) = \frac{1}{\sqrt{s}} \psi\left(\frac{t-l}{s}\right).
\end{align}
Many well known wavelets and their properties can be found in \cite{mallat1999wavelet} and references therein. A systematic method to construct wavelet families with desirable properties is presented in \cite{daubechies1992ten}. For general guidelines concerning the choice of a wavelet family, we refer to \cite{goel1995wavelet} and \cite{mallat1999wavelet}. This challenge is tackled in a geophysical context by \cite{torrence1998practical} and \cite{lovejoy2012haar}, who discuss the relation between wavelets, fluctuations and structure functions.

The continuous wavelet transform (CWT) of the discrete signal $X$ is defined as the convolution of $X$ with scaled and shifted versions of $\psi$:
\begin{align}
	CWT_X(s,l) &:= \Delta T \sum_{k=1}^{N} x_k\psi_{s,l\Delta T}(k\Delta T),
\end{align}
with scaling parameter $s>0$ and shift parameter $l\in\mathbb{R}$. The inverse transformation is given by the \textit{resolution of the identity} formula
\begin{align}
	X &= C_\psi \int_{-\infty}^{\infty} \int_{0}^{\infty} \frac{1}{s^2}\ (CWT_X)(s,l)\ \psi_{s,l}\ ds\ dl.
\end{align}
The existence of an inverse transformation is not surprising, since we do not lose any information by describing a one-dimensional function, signal or time series by the two dimensional function $CWT_X(s,l)$. This question gains significance in the discrete case, where only a finite number of scaling and shift parameters are studied.

For the discrete wavelet transform (DWT) the scaling parameter $s$ is replaced by a discrete exponential series $\{s_0^m\}_{m\in\mathbb{Z}}$ with a dilatation parameter $s_0>1$. The by far most common choice $s_0=2$ leads to dyadic decompositions. The shift parameter $l$ is discretized by the scale-dependent linear series $\{nl_0\ s_0^m\}_{n\in\mathbb{Z}}$, $l_0>0$. With
\begin{align}
	\psi^{m,n}(t) = s_0^{-m/2} \psi\left(s_0^{-m}t - nl_0\right),
\end{align}
the DWT of $X$ is given by
\begin{align}
	DWT_X(m,n) &:= \Delta T \sum_{k=1}^{N} x_k\psi^{m,n\Delta T}(k\Delta T).
\end{align}
There is no general discrete counterpart to the resolution of the identity of CWT, because information is possibly lost in the discretization step of DWT. However, we can approximate any (integrable) function $f$ by a series of wavelets to arbitrarily fine degree (\cite{daubechies1992ten}). For discrete time series we can construct bases of (possibly orthonormal) wavelets, which allow a unique (and therefore invertible) wavelet representation of $X$, e.g. via a \textit{multiresolution analysis} (MRA). In this case no information is lost in the DWT, i.e. a loss-free inverse DWT exists.

Before we discuss the MRA approach, let us compare wavelet decompositions with Fourier and Windowed-Fourier methods. Fig.~\ref{fig:time-freq-plane} shows the resolution capability of the time series and various transformations on the time-frequency plane. The original time series (a) resolves only temporal information, while its Fourier transform (b) solely resolves the frequency dimension. A time-localized frequency analysis is possible with the Windowed Fourier transform (c), which decomposes the time-frequency plane in boxes of constant shape and area. The adaptive nature of the Wavelet transform (d) is clearly visible in the changing shape of these boxes: we have a high frequency resolution, but low time resolution for low frequencies and a high time resolution but low frequency resolution for high frequencies. Note that all boxes have the same constant area. This is a direct consequence of the Heisenberg-Gabor-limit \citep{heisenberg1927anschaulichen,cohen1995time}, which states that one cannot simultaneously sharply localize a signal in both time and frequency. Through the choice of different wavelet families the wavelet transform is a very flexible tool to customize the decomposition of the time-frequency plane based on the problem at hand. A more elaborate comparison between wavelet transforms and (Windowed) Fourier transforms can be found in \cite{daubechies1992ten} and \cite{mallat1999wavelet}.

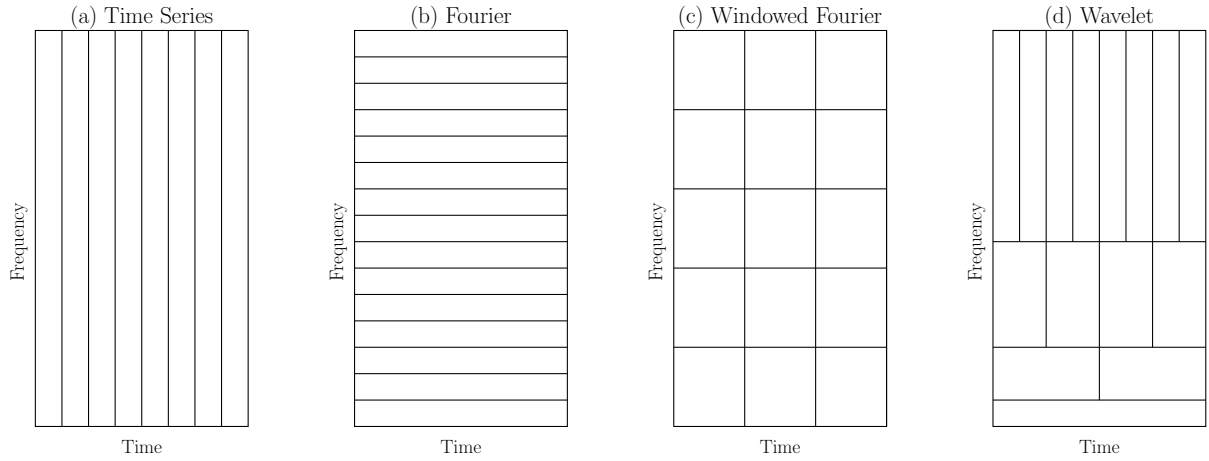
\begin{figure*}%
\centering
\begin{tikzpicture}[%
    >=stealth,
    auto,
    scale=0.35, every node/.style={transform shape}
  ]
		\def\s{12}
		\def\vs{-12}		
	\foreach \i in {0,...,3}{
	  \node (time) at (4+\s*\i,-.75) {\huge Time};
	  \node[label={[text depth=-1ex,rotate=90]center:\huge Frequency}] at (-.75+\s*\i,7) {};
	}
	  \node (a) at (4+\s*0,15.5) {\Huge (a) Time Series};
	  \node (b) at (4+\s*1,15.5) {\Huge (b) Fourier};
	  \node (c) at (4+\s*2,15.5) {\Huge (c) Windowed Fourier};
	  \node (d) at (4+\s*3,15.5) {\Huge (d) Wavelet};

	\draw[-] (0,0) rectangle (8,15);
	\foreach \i in {1,...,7}{
	  \draw[-] (\i,0) -- (\i,15);
	}

	\draw[-] (\s,0) rectangle (8+\s,15);
	\foreach \i in {1,...,14}{
	  \draw[-] (\s,\i) -- (8+\s,\i);
	}

	  \def\k{2}
	\draw[-] (\k*\s,0) rectangle (8+\k*\s,15);
	  \foreach \i in {1,...,4}{\draw[-] (\k*\s,3*\i) -- (8+\k*\s,3*\i);}
	  \foreach \i in {1,...,3}{\draw[-] (8/3*\i+\k*\s,0) -- (8/3*\i+\k*\s,15);}

	  \def\k{3}
	\draw[-] (\k*\s,0) rectangle (8+\k*\s,15);
	  \draw[-] (\k*\s,1) -- (8+\k*\s,1);
	  \draw[-] (\k*\s,3) -- (8+\k*\s,3);
	  \draw[-] (\k*\s,7) -- (8+\k*\s,7);
	  \draw[-] (\k*\s,15) -- (8+\k*\s,15);
	  \draw[-] (4+\k*\s,1) -- (4+\k*\s,3);
	  \foreach \i in {1,...,3}{\draw[-] (2*\i+\k*\s,3) -- (2*\i+\k*\s,7);}
	  \foreach \i in {1,...,7}{\draw[-] (\i+\k*\s,7) -- (\i+\k*\s,15);}

\end{tikzpicture}
\caption{The representation of a time series on the time-frequency plane is plotted for the original signal (a), and its Fourier (b), Windowed Fourier (c) and dyadic Wavelet (d) Transforms.}%
\label{fig:time-freq-plane}%
\end{figure*}

The MRA was first introduced by \cite{mallat1989theory} and successively decomposes a signal from its finest scales down to the lowest possible resolution, i.e. a globally smoothed value. Assume that we have an orthonormal family of wavelets $\psi^{m,n}$ and a function $\phi$ with
\begin{align}
	\int_{\mathbb{R}} \phi(t) dt = 1,
\end{align}
which spawns an orthonormal basis of the signal-space by translation, i.e.
\begin{align}
	\int_\mathbb{R} \phi(t-n) \phi(t-m) dt = \delta_{n,m}.
\end{align} 
For the time series $X$, the signal space is defined by functions, which are piecewise constant over an interval of length $\Delta T$. If $\phi$ relates to the mother wavelet $\psi$ via the \textit{wavelet equation} (\cite{daubechies1992ten})
\begin{align}
	\psi(t) = 2\sum_n (-1)^n \phi(2t-n) \int \phi(s)\phi(2s-n) ds,
\end{align}
it is called the \textit{scaling function} or \textit{father wavelet}. Figure \ref{fig:sf_mw} shows some widely used pairs of scaling functions and mother wavelets. In MRA, the time series is decomposed into an approximation $L = \sum_{k=1}^N x_k \phi(k\Delta T)$ and \textit{detail coefficients} given by the discrete wavelet transform $DWT_X$ with the mother wavelet. In terms of filtering the scaling function represents a low-pass filter, whereas wavelets represent high-pass filters \citep{jensen2001ripples}. The approximation $L$ can again be decomposed into an approximation $L_2$ of scale $2$ and a second set of detail coefficients $DWT_{X,2}$. Iterative application of this procedure leads to an approximation of scale $m_0$ and $m_0$ sets of wavelet coefficients. An intuitive example is the Haar wavelet (see Fig.~\ref{fig:sf_mw}). Its scaling function takes the mean-value of two neighboring data points, i.e. reducing the resolution by a factor of $2$, whereas the Haar wavelets take the difference of two neighboring data points and therefore describes differences on each scale. The MRA scheme (Fig.~\ref{fig:mra}) describes a time series of length $2^J$ by a lower-resolution \textit{approximation} of length $2^{J-1}$ and the \textit{details} of scale $J$ in form of a vector of wavelet coefficients of length $2^{J-1}$. In the second step the same procedure is applied on the lower-resolution approximation of length $2^{J-1}$, which leads to an approximation and details for scale $J-1$ of length $2^{J-2}$. Successive application of this procedure finally leads to an approximation of the lowest possible resolution, i.e. a single number, and $J$ blocks of wavelet coefficients of size $1,2,2^2,...,2^{J-1}$ describing the details on each dyadic scale.

\begin{figure}
\center
\includegraphics[width=0.5\textwidth]{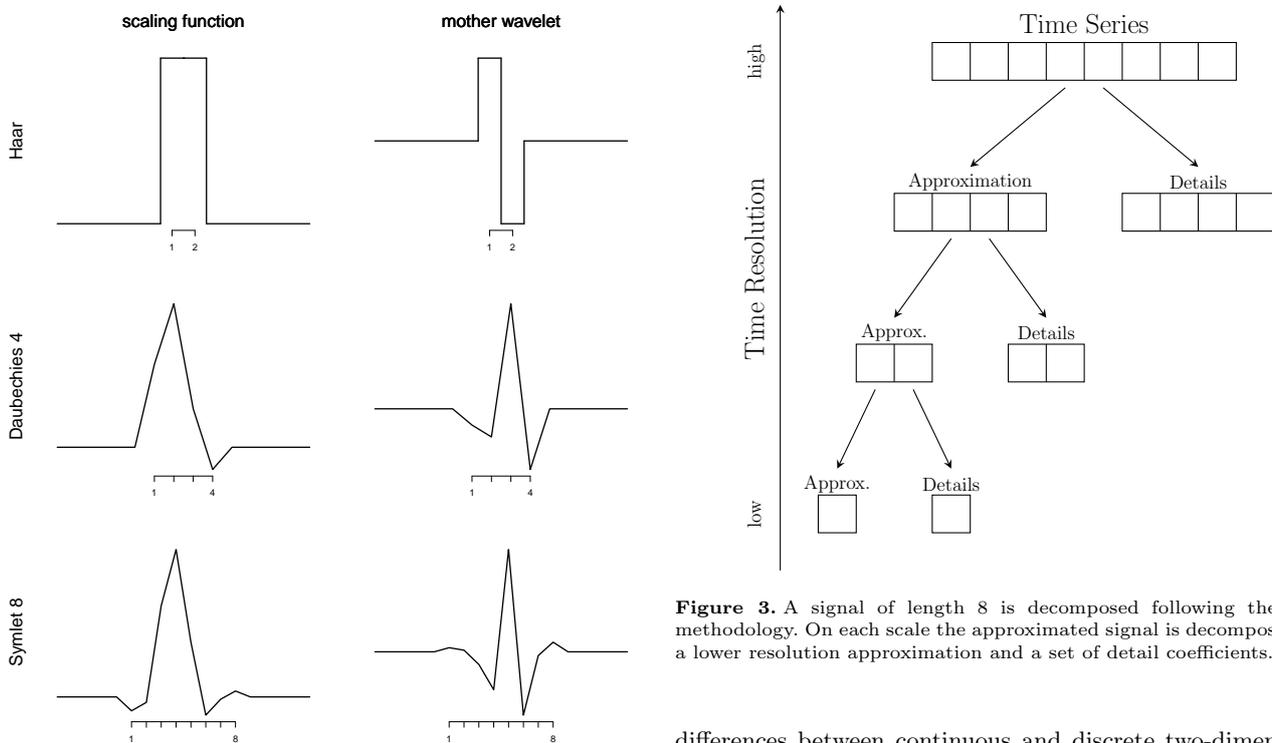}
  \caption{Discrete scaling function and mother wavelet are shown for the three widely used wavelet families. The numbers on the abscissa indicate non-zero coefficients.}
  \label{fig:sf_mw}
\end{figure}

\begin{figure}%
\centering
\begin{tikzpicture}[%
    >=stealth,
    auto,
    scale=0.5, every node/.style={transform shape}
  ]
		\def\s{2}
		\def\vs{4}	
		\def\vp{.7}	
		\draw[-] (0,0) rectangle (8,-1);
			\foreach \i in {1,...,7}{\draw[-] (\i,0) -- (\i,-1);}
		\draw[->] (3.5,-1-.2) -- (1,-\vs+\vp);
		\draw[->] (4.5,-1-.2) -- (7,-\vs+\vp);
		\node (A0) at (4,.5) {\huge Time Series};
		\draw[-] (-1,0-\vs) rectangle (3,-1-\vs);
			\foreach \i in {0,...,2}{\draw[-] (\i,0-\vs) -- (\i,-1-\vs);}
		\draw[-] (5,0-\vs) rectangle (9,-1-\vs);
			\foreach \i in {6,...,8}{\draw[-] (\i,0-\vs) -- (\i,-1-\vs);}
		\draw[->] (0.5,-1-\vs-.2) -- (-1,-2*\vs+\vp);
		\draw[->] (1.5,-1-\vs-.2) -- (3,-2*\vs+\vp);
		\node (A1) at (1,.3-\vs) {\Large Approximation};
		\node (D1) at (7,.3-\vs) {\Large Details};
		\def\k{2}
		\draw[-] (-\k,0-\k*\vs) rectangle (-\k+2,-1-\k*\vs);
			\draw[-] (-\k+1,0-\k*\vs) -- (-\k+1,-1-\k*\vs);
		\draw[-] (-\k+4,0-\k*\vs) rectangle (-\k+6,-1-\k*\vs);
			\draw[-] (-\k+5,0-\k*\vs) -- (-\k+5,-1-\k*\vs);
		\draw[->] (-1.5,-1-2*\vs-.2) -- (-2.5,-3*\vs+\vp);
		\draw[->] (-0.5,-1-2*\vs-.2) -- (0.5,-3*\vs+\vp);
		\node (A2) at (-\k+1,.3-\k*\vs) {\Large Approx.};
		\node (D2) at (-\k+5,.3-\k*\vs) {\Large Details};
		\def\k{3}
		\draw[-] (-\k,0-\k*\vs) rectangle (-\k+1,-1-\k*\vs);
		\draw[-] (-\k+3,0-\k*\vs) rectangle (-\k+4,-1-\k*\vs);
		\node (A3) at (-\k+.5,.3-\k*\vs) {\Large Approx.};
		\node (D3) at (-\k+3.5,.3-\k*\vs) {\Large Details};
		\draw[->] (-\k-1,-1-\k*\vs-1) -- (-\k-1,1);
		\node[label={[text depth=-1ex,rotate=90]center:\huge Time Resolution}] at (-\k-1.75,-1.5*\vs) {};
		\node[label={[text depth=-1ex,rotate=90]center:\Large low}] at (-\k-1.75,-3*\vs-.5) {};
		\node[label={[text depth=-1ex,rotate=90]center:\Large high}] at (-\k-1.75,-.5) {};
\end{tikzpicture}
\caption{A signal of length $8$ is decomposed following the MRA methodology. On each scale the approximated signal is decomposed into a lower resolution approximation and a set of detail coefficients.}%
\label{fig:mra}%
\end{figure}

\subsection{Two-Dimensional Wavelet Transforms}
\label{subsec:2d-wavelet-transform}
There are many different possibilities to extend wavelet transforms to two dimensions. We present the most common approaches used in spatial verification. Since there are subtle differences between continuous and discrete two-dimensional transforms, these cases will be discussed separately. In this section we assume that the data to be analyzed are given in the form of a matrix $X=(x_{ij})_{i,j=1}^N$.

The natural way to extend the CWT to two dimensions is to replace the shift parameter by a two dimensional vector $\mathbf{l}=(l_x,l_y)\in\mathbb{R}^2$. However, there are now even more possibilities to choose a mother wavelet. The straightforward option is to use the rotation body of a one-dimensional wavelet, i.e. using functions only depending on absolute values of location and shift parameters. The rotational invariant or \textit{non-directional} 2D-CWT is then given by
\begin{align}
	CWT_X(s,\mathbf{l}) &:= \frac{1}{s} \sum_{i=1}^{N}\sum_{j=1}^{N} x_{ij}\psi\left(\frac{\left\|(i,j)-\mathbf{l}\right\|}{s}\right).
\end{align}
Note that it is sometimes convenient to replace the scale-normalization factor $\frac{1}{s}$ by $s^{-2}$. The former leads to a conservation of energy across different scales due to the identity
\begin{align}
	\int_{\mathbb{R}^2} \left| \frac{1}{s}\psi\left(\frac{\left\|\mathbf{r}-\mathbf{l}\right\|}{s}\right) \right|^2 d\mathbf{r} = \int_{\mathbb{R}^2} \left| \frac{1}{s'}\psi\left(\frac{\left\|\mathbf{r}-\mathbf{l'}\right\|}{s'}\right) \right|^2 d\mathbf{r},
\end{align}
for all $s,s'>0$ and $\mathbf{l},\mathbf{l}'\in\mathbb{R}^2$. The latter leads to an analogous identity in the $L^1$-norm (i.e. the integral over absolute values) see \cite{dallard1993cwt} for more details.

It is also possible to add a directional component to the 2D-CWT by choosing a wavelet that is sensitive to rotation. This adds an additional parameter, the rotation angle $\Theta\in[0,2\pi]$, and leads to the \textit{directional} 2D-CWT
\begin{align}
	CWT_X(s,\mathbf{l},\Theta) &:= \frac{1}{s} \sum_{i=1}^{N}\sum_{j=1}^{N} x_{ij}\psi\left(R_\Theta^{-1}\left(\frac{(i,j)-\mathbf{l}}{s}\right)\right),
\end{align}
where $R_\Theta$ is the standard two-dimensional rotation matrix of angle $\Theta$. A comprehensive comparison between directional and non-directional 2D CWT with applications to meteorological data can be found in \cite{wang2010two}.

In the discrete case the most common approach to two-dimensional data is a two-time application of the one-dimensional DWT, first on the rows of the data matrix X, then on columns. This leads to the window-scheme \ref{fig:window-scheme}, which decomposes an $N\times N$ matrix into four $N/2 \times N/2$ blocks: vertical, horizontal and diagonal wavelets coefficients and a smoothed approximation of the original data. Analogous to the one-dimensional MRA, only the approximation-block is further decomposed for successive DWT-levels. It is possible to define other schemes, e.g. leading to non-directional decompositions via lifting schemes, see for instance \cite{jensen2001ripples}.

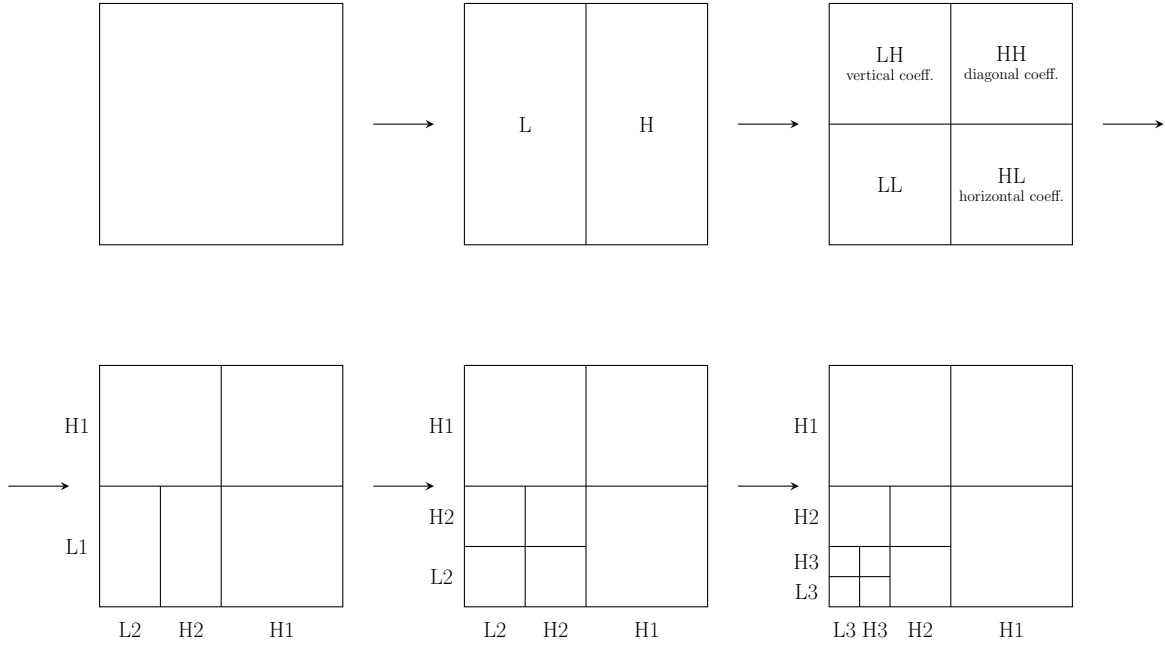
\begin{figure*}%
\centering
\begin{tikzpicture}[%
    >=stealth,
    auto,
    scale=0.4, every node/.style={transform shape}
  ]
		\def\s{12}
		\def\vs{-12}		
	\draw[-] (0,0) rectangle (8,8);
	
	\draw[->] (8+1,4) -- (\s-1,4);
	
	\draw[-] (0+\s,0) rectangle (8+\s,8);
	\draw[-] (4+\s,0) -- (4+\s,8);
		\node (L) at (2+\s,4) {\huge L};
	\node (H) at (6+\s,4) {\huge H};
			
	\draw[->] (\s+8+1,4) -- (2*\s-1,4);
	
	\draw[-] (0+2*\s,0) rectangle (8+2*\s,8);
	\draw[-] (4+2*\s,0) -- (4+2*\s,8);
	\draw[-] (0+2*\s,4) -- (8+2*\s,4);
		\node (LL) at (2+2*\s,2) {\huge LL};
	\node[align=center] (LH) at (2+2*\s,6) {{\huge LH}\\[.2cm] \Large{vertical coeff.}};
	\node[align=center] (HL) at (6+2*\s,2) {{\huge HL}\\[.2cm] \Large{horizontal coeff.}};
	\node[align=center] (HH) at (6+2*\s,6) {{\huge HH}\\[.2cm] \Large{diagonal coeff.}};
	
	\draw[->] (2*\s+8+1,4) -- (3*\s-1,4);
	
	\draw[->] (8+1-\s,4+\vs) -- (\s-1-\s,4+\vs);
	
	\draw[-] (0,0+\vs) rectangle (8,8+\vs);
	\draw[-] (4,0+\vs) -- (4,8+\vs);
	\draw[-] (0,4+\vs) -- (8,4+\vs);
	\draw[-] (2,0+\vs) -- (2,4+\vs);
	\node (L1_left) at (-.75+0*\s,2+\vs) {\huge L1};
	\node (H1_left) at (-.75+0*\s,6+\vs) {\huge H1};
	\node (L2_bot) at (1+0*\s,-.75+\vs) {\huge L2};
	\node (H2_bot) at (3+0*\s,-.75+\vs) {\huge H2};
	\node (H1_bot) at (6+0*\s,-.75+\vs) {\huge H1};

	\draw[->] (8+1,4+\vs) -- (\s-1,4+\vs);
	
	\draw[-] (0+\s,0+\vs) rectangle (8+\s,8+\vs);
	\draw[-] (4+\s,0+\vs) -- (4+\s,8+\vs);
	\draw[-] (0+\s,4+\vs) -- (8+\s,4+\vs);
	\draw[-] (0+\s,2+\vs) -- (4+\s,2+\vs);
	\draw[-] (2+\s,0+\vs) -- (2+\s,4+\vs);
	\node (L2_left) at (-.75+1*\s,1+\vs ) {\huge L2};
	\node (H2_left) at (-.75+1*\s,3+\vs ) {\huge H2};
	\node (H1_left) at (-.75+1*\s,6+\vs ) {\huge H1};
	\node (L2_bot)  at ( 1+1*\s,-.75+\vs) {\huge L2};
	\node (H2_bot)  at ( 3+1*\s,-.75+\vs) {\huge H2};
	\node (H1_bot)  at ( 6+1*\s,-.75+\vs) {\huge H1};
	
	\draw[->] (\s+8+1,4+\vs) -- (2*\s-1,4+\vs);
	
	\draw[-] (0+2*\s,0+\vs) rectangle (8+2*\s,8+\vs);
	\draw[-] (4+2*\s,0+\vs) -- (4+2*\s,8+\vs);
	\draw[-] (0+2*\s,4+\vs) -- (8+2*\s,4+\vs);
	\draw[-] (0+2*\s,2+\vs) -- (4+2*\s,2+\vs);
	\draw[-] (2+2*\s,0+\vs) -- (2+2*\s,4+\vs);
	\draw[-] (0+2*\s,1+\vs) -- (2+2*\s,1+\vs);
	\draw[-] (1+2*\s,0+\vs) -- (1+2*\s,2+\vs);
	\node (L3_left) at (-.75+2*\s,.5+\vs) {\huge L3};
	\node (H3_left) at (-.75+2*\s,1.5+\vs){\huge H3};
	\node (H2_left) at (-.75+2*\s,3+\vs ) {\huge H2};
	\node (H1_left) at (-.75+2*\s,6+\vs ) {\huge H1};
	\node (L3_bot)  at (.5+2*\s,-.75+\vs) {\huge L3};
	\node (H3_bot)  at (1.5+2*\s,-.75+\vs){\huge H3};
	\node (H2_bot)  at ( 3+2*\s,-.75+\vs) {\huge H2};
	\node (H1_bot)  at ( 6+2*\s,-.75+\vs) {\huge H1};
\end{tikzpicture}
\caption{The window-scheme for a 2D MRA is shown. In the first step, the first level of a 1D MRA is applied to the rows of the original image. The second step applies the same MRA on the columns yielding 4 distinct sub-images: one low-resolution approximation (LL) and three detail coefficients (vertical (LH), horizontal (HL) and diagonal (HH)). As in the one-dimensional MRA these steps are iterated on the low-resolution approximation to derive higher level decompositions.}%
\label{fig:window-scheme}%
\end{figure*}

The MRA is an example of a \textit{decimated} DWT, i.e. the number of wavelet coefficients is equal to the number of original data points. A weakness of the MRA is its shift sensitivity and poor location information on larger scales \citep{mallat1999wavelet}.  \textit{Redundant} or \textit{non-decimated} DWT use additional wavelet coefficients to tackle this issue. These methods are known as \textit{frame expansions} in mathematical literature \citep{daubechies1992ten}. The standard \textit{Redundant Discrete Wavelet Transform} (RDWT), also known as \textit{Algorithme \`{a} trous}, calculates the detail coefficients for each scale and each of the three directions (vertical, horizontal, diagonal) for every original data point. The RDWT represents information of any scale and direction at the full original resolution, is shift-invariant and robust to noise \citep{fowler2005redundant}. However, this comes at the price of high redundancy: for a RDWT of level $m$, a signal of length $N$ is represented by a wavelet frame with $3m\times N$ wavelet coefficients. Another popular way to deal with shift sensitivity is the \textit{Dual-Tree Complex Wavelet Transform} (DTCWT), which uses two separate, phase-shifted MRA trees to represent the original signal \citep{selesnick2005dual,kingsbury2001complex}.

\section{Wavelet Based Spatial Verification Techniques}
\label{sec:spatial-verification-techniques}

\subsection{Scale Separation and Point-Measure Enhancement}
\label{subsec:scale-separation-and-point-measures}

While \cite{kumar1993multicomponent} do not explicitly tackle the issue of spatial verification, they pioneered the application of two-dimensional MRA for quantitative precipitation forecasts (QPF), which is the cornerstone for many, if not most, of the wavelet based spatial verification techniques used today. They study the hypothesis that rainfall, analogous to synthetic stochastic processes, can be decomposed in a large-scale component representing the mean behavior of the process, and a small scale component consisting of self-similar fluctuations. A MRA (i.e. non-redundant 2D DWT) is applied to two-dimensional precipitation fields. The final level of approximation, i.e. the lowest resolution $m_0$, is determined from statistical considerations as the largest scale where rainfall fluctuations still exhibit self-similarity. The precipitation process $X$ is decomposed into
\begin{align}
		X &= \bar{X} + \sum_{m\geq m_0} X'_m \\
		X' &= X'_{1,m} + X'_{2,m} + X'_{3,m},
\end{align}
where $\bar{X}$ denotes the low-resolution approximation and $X'_{1,m}$, $X'_{2,m}$, $X'_{3,m}$ the coefficients of the two-dimensional DWT in vertical, horizontal and diagonal direction (see Fig.~\ref{fig:window-scheme}). The results from statistical analysis of the small scales can then for instance be used to improve (directional) sub-grid parametrization, e.g. for convective precipitation. Based on these premisses, \cite{perica1996linkage,perica1996model} developed a statistical downscaling method for mesoscale precipitation forecasts. \cite{benedetti2005verification} used this method as a tool to verify ECMWF forecasts for rainfall in tropical cyclones against radar observations. There, it is applied in the opposite direction, to reconstruct a statistical average from high-resolution measurements.

\cite{briggs1997wavelets} use two-dimensional DWTs on $500$hPa-height fields with two goals in mind. First, to derive a method to improve the standard (pointwise) forecast performance scores root mean square error (RMSE) and anomaly correlation coefficient (ACC) \citep{miyakoda1972cumulative} by removing insignificant information. This is also known as \textit{denoising}, see for instance \cite{mallat1999wavelet}. Second, they develop multivariate measures, which aim at providing more information on the closeness of two fields, than point scores. The first step, however, is to find a suitable mother wavelet for the data at hand. To this end several DWTs with different mother wavelets are carried out. Following \cite{goel1995wavelet} the DWT with the smallest Shannon entropy measure
\begin{align}
	H(DWT) = -\sum_{ij}w'_{ij} \log{w'_{ij}},
\end{align}
where $w'_{ij}$ are the non-negative normalized wavelet coefficients, yields the best separation of scales. After the data are transformed to wavelet space, various thresholding techniques are applied to the wavelet coefficients. \cite{briggs1997wavelets} conclude that hard thresholding of the form 
\begin{align}
		w_{ij} = 	\begin{cases}
								0&, |w_{ij}| < \lambda \\
								w_{ij}&, |w_{ij}| \geq \lambda
							\end{cases},
\end{align}
with a positive constant $\lambda$ is optimal for data compression and soft thresholding of the form
\begin{align}
		w_{ij} = 	\text{sign}(w_{ij})\max\left(0,\ |w_{ij}|-\lambda\right)
\end{align}
is preferable in statistical analysis settings. After thresholding, the inverse DWT is carried out, and the resulting denoised spatial fields are evaluated with the pointwise measures, RMSE and ACC, which both improved. However, we would like to emphasize that denoising removes weak (small-scale) variability. Whether or not this variability can be considered as noise (i.e. not significant for the analysis of the data at hand) has to be carefully evaluated by the user and cannot be decided by the thresholding techniques. To derive a multivariate measure of closeness, two-dimensional MRA is applied to orthogonally decompose the spatial fields into dyadic scales. The MRA scheme used by \cite{briggs1997wavelets} differs from the window-scheme presented in Fig.\ref{fig:window-scheme} as they apply the transformation first in one direction for all scales then in the other. This leads to 2D wavelets where the x and y component might have different scales. For each scale, RMSE and ACC are derived separately and their percent contributions to the overall error scores are calculated, to allow an analysis of the fields at distinct (dyadic) spatial scales. \cite{briggs1997wavelets} further show that both RMSE and ACC can be calculated in data space (following the above procedure) or wavelet space, i.e. directly using the coefficients of the wavelet transform. In the latter case the number of coefficients $N^2_l$ varies and, denoting the total number of grid points in data space by $N^2$, the following relation holds true
\begin{align}
	\frac{N}{N_l} \text{RMSE}_l(data) &= \text{RMSE}_l(wavelet)
\end{align}
for each spatial scale $l = 1, \ldots, L$. We denote verification approaches that follow the methodology by \cite{kumar1993multicomponent} and \cite{briggs1997wavelets} as \textit{Point-Measure Enhancement} (PME) techniques.

\subsection{Intensity-Scale Skill Score}
\label{subsec:intensity-scale-skill-score}
The most popular wavelet-based spatial verification method to date is certainly the intensity-scale skill score (ISS) introduced by \cite{casati2004wavelet}. It is based on the work of \cite{briggs1997wavelets} and uses MRA with traditional pointwise measures to assess the quality of precipitation on each scale, separately. However, it extends this approach by data pre-processing and -recalibration, which allows the scale separation of widely used (categorical) skill scores. The ISS methodology follows five steps:
\newpage
\begin{enumerate}
	\item data pre-processing: dithering
	\item forecast recalibration
	\item binary masks (thresholding)
	\item binary error decomposition
	\item calculation of skill scores
\end{enumerate}
\cite{casati2004wavelet} use $256\times 256$ forecast and analysis precipitation rate fields. In the first step, all non-zero precipitation are \textit{dithered}, i.e. a small amount of uniformly distributed noise in the range of $(-1/64,1/64)mm/h$ is added to the data to reduce the discretization round-off error. Forecasts $Y$ are then recalibrated, so that their empirical univariate cumulative distribution function (CDF) $F_Y$ matches the CDF of the analysis $F_X$:
\begin{align}
	Y' = F_X^{-1}\left( F_Y(Y) \right).
\end{align}
This non-linear transformation eliminates (and implicitly gives a measure of) the bias in the marginal distributions of the forecast. Thresholding over a number of dyadic precipitation rates $0,2^{-5},\ldots,2^{7}$ converts the recalibrated forecast and the analysis into binary images $I_{Y'}, I_X$. The binary error is defined as the difference of those images
\begin{align}
	Z = I_{Y'} - I_X.
\end{align}
The binary error, which is a $256\times 256$ matrix is decomposed into $L=8$ different scales $Z_l$ using a two-dimensional DWT with the Haar wavelet, which leads to
\begin{align}
	Z = \sum_l^L Z_l.
\end{align}
Due to the orthonormality of the Haar wavelet family the MSE of $Z$ can be expressed as a sum over the MSE of the separate scales, i.e.
\begin{align}
	MSE(Z) = \sum_{l=1}^L MSE(Z_l).
\end{align}
A skill score is defined with a random forecast as reference, taking into account the base rate $\epsilon$, i.e. the fraction of ones in the binary analysis image $I_X$:
\begin{align}
	SS(Z) = \frac{MSE(\text{random})-MSE(Z)}{MSE(\text{random})} = 1 - \frac{MSE(Z)}{2\epsilon (1-\epsilon)}.
\end{align}
$2\epsilon(1-\epsilon)$ is the expected value of a random forecast with no spatial correlations and base rate $\epsilon$. Skill scores on a particular scale $l=1,\ldots,L$ are defined by
\begin{align}
	SS_l(Z) = 1 - \frac{MSE(Z_l)}{2\epsilon (1-\epsilon) / L},
\end{align}
which leads to the skill score decomposition
\begin{align}
	SS(Z) = \frac{1}{L}\sum_{l=1}^L SS_l(Z).
\end{align}
Including the base rate as a weight for the MSE compensates for the double penalty of displacement errors prevalent in pointwise scores (see for instance \cite{gilleland2009}). The binary character of the transformed fields allows to bridge the gap to traditional categorical scores, i.e. Heidke skill score (HSS), \cite{heidke1926berechnung}, and Pierce skill score (PSS), \cite{peirce1884numerical,peirce1993writings}. \cite{casati2004wavelet} show that, for the pre-processed binary data, all three skill scores are equivalent
\begin{align}
	SS = HSS = PSS.
\end{align}
Figure \ref{fig:iss} shows the ISS plotted as a two-dimensional function of scale and threshold. This enables the user to derive information regarding skillful scales and to pinpoint different synoptic events as culprits of low skill such as missed showers or displaced fronts. The ISS is part of the \textit{Model Evaluation Tool} (MET), \cite{brown2009model}, the R-package for spatial verification \textit{SpatialVx} \citep{spatialvx} and the Land surface Verification Toolkit (LVT), \cite{kumar2012land}.

\cite{casati2007new} extend the methodology of the ISS to probabilistic forecasts. Forecasts and observations of probabilities of lightning occurrence from the Canadian Meteorological Centre (CMC) are evaluated. Two-dimensional MRA is used to decompose a field $X$ into detail- and approximation-fields. The squared energy of a field $X=(x_{ij})_{i,j=1}^N$ is used as a measure of overall quantity of events
\begin{align}
	E = \frac{1}{N^2}\sum_{i,j=1}^N x_{ij}^2.
\end{align}
Due to the orthonormality of the Haar wavelet family, the MRA decomposition can be used to calculate the squared energy contribution of each scale. Dominant scales are associated to peaks in the energy spectrum. \cite{casati2007new} use a scale decomposition of the Brier score (BS), \cite{Brier50}, analogous to the ISS case to derive skillful scales for different lightning events. A very appealing characteristic of this approach is its compatibility with the classical Brier score decomposition (\cite{Murphy73a})
\begin{align}
	BS = \text{reliability} - \text{resolution} + \text{uncertainty},
\end{align}
i.e. the BS decomposes into reliability, resolution and uncertainty on each separate spatial scale. This allows for a detailed scale-resolving evaluation of the probabilistic characteristics of the forecast. 

In a revision of the ISS technique, \cite{casati2010new} omitted dithering and (more importantly) recalibration of the forecast fields. Therefore, bias has to be taken into account. For the whole field this is accomplished by looking at the lowest resolution scale, where the whole domain is smoothed to a single grid point, see Fig.~\ref{fig:mra}. In case of the Haar wavelet this corresponds to a field-wide mean value. To assess the bias on particular scales, the energy of the thresholded, and therefore binary, fields is analyzed analogously to \cite{casati2007new}. The energy relative differences are compared between forecast (F) and observation (O) on each scale
\begin{align}
	E_{rel.diff} = \frac{E(F)-E(O)}{E(F)+E(O)}.
\end{align}
In addition, the percent contributions of each scale to the total $E_{rel.diff}$ are studied to analyze the scale-structures of forecast and observation. Note that, contrary to the ISS, energy-measures as well as MSE values depend on the number of events.
Four different ways to deal with non-dyadic field dimensions are discussed: \textit{padding} (adding constant values until the next dyadic dimension is reached), \textit{cropping} (choosing a dyadic sub-domain), \textit{interpolation} onto a finer grid and \textit{tiling} (moving windows of dyadic size are used to calculate several realizations of skill scores). In addition to these techniques, the R-package \textit{SpatialVx} is able to employ the \textit{maximal overlap method} (MODWT) via its dependency package \textit{waveslim} \citep{waveslim}. MODWT is able to handle data with non-dyadic dimensions and is less sensitive to translation errors, but computationally more expensive than traditional MRA algorithms. We refer to \cite{percival1994long} and \cite{liang1994two,liang1997image} for further details.
Geometric test cases and synthetically disturbed forecasts from the spatial verification intercomparison project (ICP), \cite{gilleland2009}, show sensitivity of ISS towards errors in bias and in size and location of features. The latter is a direct consequence of the non-redundant MRA decomposition. In order to avoid confusion in the discussion of application-papers, we will denote the revised ISS by \ISS. 

\cite{saux2012wavelet} extend the ISS method in two ways. First, data dependent thresholds are introduced based on quantiles of the forecast and observation field, respectively (e.g. 0\%-quantile ($q1$), 20\%-quantile ($q2$), $\ldots$, 80\%-quantile ($q5$), 100\%-quantile ($q6$)). The binary images are calculated based on intervals created by these  quantiles, e.g.
\begin{align}
\label{eq:saux_ix}
		I_X = 	\begin{cases}
								1, &q1(X) \leq X < q2(X) \\
								0, &else
						\end{cases}
\end{align}
and
\begin{align}
\label{eq:saux_iy}
		I_Y = 	\begin{cases}
								1, &q1(Y) \leq Y < q2(Y) \\
								0, &else
						\end{cases}
\end{align}
In general $q1(X) \neq q1(Y)$, i.e. different absolute values (e.g. of precipitation) are compared with each other. The binary error image for a quantile $q$ is denoted by $Z_q=I_Y-I_X$. This procedure is closely connected to the recalibration step of \cite{casati2004wavelet}, since the recalibrated forecast
\begin{align}
	Y' = F_X^{-1}\left( F_Y(Y) \right)
\end{align}
and the observations $X$ have equal quantiles (omitting the effect of point masses for $\{Y=0\}$). Note that binary maps in \eqref{eq:saux_ix} and \eqref{eq:saux_iy} are not defined as threshold exceedances, but on an interval. The second extension to the ISS methodology tackles the issue of missing data. A binary \textit{weight image}
\begin{align}
\label{eq:saux_weight_image}
		\zeta_0 = 	\begin{cases}
								1, &\text{for valid data} \\
								0, &\text{for missing data}
						\end{cases}
\end{align}
is defined and introduced into the wavelet decomposition of the binary difference image $Z_q$ as a weight in the smoothing function. Since Haar wavelets are used in the DWT, the smoothing function $\phi_l$ of scale $l$ is just the spatial average over an area of $2^l\times 2^l$ gridpoints. The weighted smoothing function is defined as
\begin{align}
\label{eq:saux_weight_wavelet}
		\phi_l(Z_q) = \frac{\langle Z_q \zeta_0 \rangle_{2^l\times 2^l}}{\langle \zeta_0 \rangle_{2^l\times 2^l}},
\end{align}
where $\langle \cdot \rangle_{2^l\times 2^l}$ denotes the spatial average over a $2^l\times 2^l$ domain. The detail coefficients are calculated as the difference between two approximations of neighboring scales
\begin{align}
\label{eq:saux_weight_mse}
		\psi_l(Z_q) = \phi_{l-1}(Z_q) - \phi_{l}(Z_q).
\end{align}
This way, the weight function allows for an orthogonal scale decomposition and ISS calculation of incomplete sets of data. Since the definition of binary images based on quantile intervals leads to different ISS values, we will denote this approach by \ISSq  to avoid confusion.

\subsection{Verification Measures in Wavelet Space}
\label{subsec:verification-measures-in-wavelet-space}
\cite{livina2008wavelet} propose a scalar score for the comparison of spatial climate fields based on coefficients from a two-dimensional DWT. It originates from the reduction of variance skill score or \textit{Nash-Sutcliffe coefficient} \citep{nash1970river}, i.e. the MSE skill score with sample climatology $\overline{X}$ as references
\begin{align}
\label{eq:ns-coeff}
		NSC := 1 - \frac{\sum_i (Y_i - X_i)^2}{\sum_i (X_i - \overline{X})^2},
\end{align}
for forecast $Y$ and observation $X$. \cite{livina2008wavelet} decompose the spatial fields with a two-dimensional MRA following the window-scheme (see Fig.~\ref{fig:window-scheme}). The fields of wavelet coefficients of scale $l=0,\ldots,m$ and direction $d$ (1$=$vertical, 2$=$horizontal, 3$=$diagonal, 4$=$low-resolution approximation) are denoted by $C^{l,k}_X$ and $C^{l,k}_Y$ for observation and forecasts, respectively. Small values of $l$ indicate small scales, whereas large numbers of $l$ represent coarser resolutions\footnote{There are some inconsistencies in \cite{livina2008wavelet} between the mathematical definitions of the wavelet decomposition and the description of scales and directions of the resulting coefficient fields. We have chosen to change descriptions and indices to follow the mathematical definitions.}. The approximation at the lowest scale is therefore given by $C^{m,4}$. To derive a wavelet based skill score similar to the Nash-Sutcliffe coefficient, a symmetric version of $NSC$ is defined by
\begin{align}
\label{eq:nsc-symm}
		NSC_{symm} := 1 - \frac{2\sum_i (Y_i - X_i)^2}{\sum_i (X_i - \overline{X})^2 + (Y_i - \overline{Y})^2}.
\end{align}
For each scale and direction this error is calculated on the fields of wavelet coefficients, and summarized in a weighted average.
{\setlength{\alignnumbersep}{-10mm}
\scalebox{.89}{\parbox{\linewidth}{%
\begin{align}
\label{eq:wcs}
		WCS :&= \frac{1}{m+1}\sum_{l=0}^m \frac{S_l + (m-l) SA_m}{(m-l+1)}\hspace*{3cm} \\
		SA_m &= 1- \frac{2\sum_{i,j}\left(C_Y^{m,4}(i,j) - C_X^{m,4}(i,j)\right)^2} {\sum_{i,j}\left(C_X^{m,4}(i,j) - \overline{C_X^{m,4}}\right)^2 + \left(C_Y^{m,4}(i,j) - \overline{C_Y^{m,4}}\right)^2} \notag \\
		S_l &= 1- \frac{2\sum_{d=1}^3\sum_{i,j}\left(C_Y^{l,d}(i,j) - C_X^{l,d}(i,j)\right)^2} {\sum_{d=1}^3\sum_{i,j}\left(C_X^{l,d}(i,j) - \overline{C_X^{l,d}}\right)^2 + \left(C_Y^{l,d}(i,j) - \overline{C_Y^{l,d}}\right)^2}, \notag
\end{align}
}}
}
where $(i,j)$ indicates the spatial position of the detail coefficient. $SA_m$ represents the skill score of the low-resolution approximation at scale $l=m$, whereas $S_l$ represents the skill of the forecast at each scale $l$, summarized over all three directions. The scale-based weights $1/(m-l+1)$ are chosen so that larger scales (i.e. large values of $l$) have a larger influence on $WCS$. The low-resolution approximation has the strongest impact due to the factor $(m-l)$. The WCS score lies in the interval $(-\infty,1]$ and equals one for a perfect forecast. \cite{livina2008wavelet} carry out sensitivity tests for different wavelet families and study the behavior of the score on synthetically perturbed data sets and case studies against traditional scores (i.e. correlation coefficient, percentage bias and RMSE). They find that the $WCS$ score is able to distinguish between climate models in agreement with visual evaluation and does so more efficiently than traditional scores in cases with significant bias. There is a methodological difference between this approach and the ones by \cite{casati2010new}, or \cite{kwiatkowski2014spatial}: the scale separation in the calculation of the ISS is focused on the image-space, since it is defined via deviations between two approximations of different resolutions. The WCS skill score uses a (directional) two-dimensional DWT to calculate scores directly on the detail coefficients (i.e. in wavelet space). While this methodological difference is attenuated by the summation over all directions in the definition of $S_l$, one could easily modify the $WCS$ to look at the skill contribution of each scale in a particular direction. We would like to emphasize that the $WCS$ is designed as a climatological score, hence the behavior on large scales clearly dominates the overall score.

\cite{turner2004predictability} and \cite{germann2006predictability} apply wavelets to study the predictability of precipitation nowcasts from radar data. A two-dimensional discrete Haar wavelet transform is used to decompose radar observations $X$ and nowcasting predictions $Y$, yielding fields of wavelet coefficients $C_X^{l,d}$ and $C_Y^{l,d}$, for scale $l$ and direction $d$ (1$=$vertical, 2$=$horizontal, 3$=$diagonal)\footnote{We have changed the notation of \cite{turner2004predictability} to be consistent with \cite{livina2008wavelet}.}. An important difference to previously discussed methods is the calculation of a \textit{redundant} 2D DWT (in contrast to the non-redundant MRA), i.e. $C_X^{l,d}$ and $C_Y^{l,d}$ have the same dimension as the original fields $X$ and $Y$ for each scale and each direction. Wavelet spectra and Co-spectra are defined as
\begin{align}
\label{eq:turner-wavelet-spectra}
		\mathcal{S}_{XX}(l) &= \left\langle 4^{-l}\sum_{d=1}^{3}\left(C_X^{l,d}\right)^2 \right\rangle \notag\\
		\mathcal{S}_{YY}(l) &= \left\langle 4^{-l}\sum_{d=1}^{3}\left(C_Y^{l,d}\right)^2 \right\rangle \\
		\mathcal{S}_{XY}(l) &= \left\langle 4^{-l}\sum_{d=1}^{3}C_X^{l,d}\ C_Y^{l,d} \right\rangle, \notag
\end{align}
where $\langle \cdot \rangle$ denotes the spatial average over the whole domain. The normalization factor $4^{-l}$ is a direct consequence of the redundant 2D DWT and corresponds to the area of the compact support of the 2D Haar wavelet at scale $l$ (see appendix A in \cite{turner2004predictability}). The wavelet decomposition then provides
\begin{align}
\label{eq:turner-wavelet-decomp}
		\langle Y^2 \rangle &= \langle Y \rangle^2 + \sum_l \mathcal{S}_{YY}(l) \\
		\langle X^2 \rangle &= \langle X \rangle^2 + \sum_l \mathcal{S}_{XX}(l), \notag
\end{align}
which allows for the following MSE decomposition
\begin{align}
\label{eq:turner-mse-decomp}
		MSE &= \langle (Y - X)^2 \rangle \\
		&= \left(\langle Y \rangle - \langle X \rangle\right)^2 + \sum_l \mathcal{S}_{YY}(l) -2\mathcal{S}_{XY}(l) + \mathcal{S}_{XX}(l). \notag
\end{align}
The first term on the right hand side corresponds to the error on the largest scale, i.e. the bias. The error for each scale can be expressed as
\begin{align}
\label{eq:turner-scale-error}
		MSE(l) = \mathcal{S}_{YY}(l) -2\mathcal{S}_{XY}(l) + \mathcal{S}_{XX}(l).
\end{align}
As in the methodology of the ISS, it is the orthogonality of the wavelet family that allows to represent the total MSE as a sum over its contributions on each scale. To improve the nowcasting, scale-dependent weights $w(l)$ are introduced, which are multiplied with the wavelet coefficients before an inverse DWT is carried out. The error of the resulting weighted forecast $Y_w$ is given by
\begin{align}
\label{eq:turner-ff-error}
		MSE_{Y_w}(l) = w(l)^2 \mathcal{S}_{YY}(l) - 2w(l)\ \mathcal{S}_{XY}(l) + \mathcal{S}_{XX}(l).
\end{align}
To reduce the nowcasting error, the weights are chosen such that they minimize the MSE on each scale and hence, due to \eqref{eq:turner-mse-decomp}, the total forecasting error
\begin{align}
\label{eq:turner-optimal-weights}
		w_{\text{optimal}}(l) = \frac{\mathcal{S}_{XY}(l)}{ \mathcal{S}_{YY}(l)}.
\end{align}
Since these require knowledge about forecast and observations, a training period is used to derive the weights, which are then scaled depending on the lead time. \cite{turner2004predictability} show that this approach significantly reduces the forecast error and improves the correlation coefficient of the nowcasts. In this methodology, the wavelet transform is not explicitly used to measure forecast error but rather to minimize it. However, it could be easily adapted for verification purposes, for instance by using the weights $w(l)$ as indicators for skillful scales.

\subsection{Two-Dimensional CWT}
\label{subsec:2d-cwt}
The first meteorological application of two dimensional continuous wavelet transforms was in turbulence studies, see e.g. \cite{farge1992wavelet} for comprehensive review. \cite{dallard1993cwt} introduced the isotropic halo and arc wavelets, whereas \cite{antoine1996two} use a directional 2D CWT with a rotation parameter to improve the ability to detect singularities in a set of data along specific directions. To the best of our knowledge there is no spatial verification score or refined methodology based on 2D CWT. However, \cite{lu2010scale,wang2010two} use 2D CWT as a diagnostic tool in the evaluation of precipitation and wind fields. \cite{lu2010scale} apply a 2D CWT with an isotropic Halo wavelet to global QPFs and satellite observations to extract distributions for two spatial scales (200km and 1000km). Those are then evaluated with scatter plots and correlation coefficients. \cite{wang2010two} give a comprehensive introduction on the application of 2D CWT to meteorological data. The mathematical foundation of 2D CWT is briefly discussed, including some common choices for the wavelet family (Morlet, Halo and Arc, Mexican Hat, Cauchy, Poisson) as well as application specific wavelets such as the Difference of Gaussian wavelet. Presented areas of application are denoising, scale separation, field reconstruction and feature localization. The latter demonstrates a merging point between spatial scale separation verification methods and feature based techniques. Particular attention is paid to the freedom of choice regarding spatial scale and directionality of the 2D CWT. This allows for the detection of coherent wave structures of a specific scale and direction in wind fields. While no rigid verification methodology is presented, the authors show the potential of 2D CWT in the context of spatial verification.

\begin{table*}[p]
\center
\begin{tabular}[ht]{p{2cm}p{1.4cm}p{1.4cm}p{2.2cm}p{4.5cm}p{1.8cm}} 
		\toprule[2pt]
		Name / Description of Method & Wavelet Transformation & Data Requirements & Results & Applications & Method-Relevant Publications \\
		\midrule[2pt]
		Scale Separation and Point-Measure Enhancement (PME) & usually MRA & usually dyadic & denoising, scale separation of traditional pointwise verification scores & \cite{brunsell2003length}, \cite{alvera2007forecast}, \cite{smith2011multiscale}, \cite{singh2011statistical}, \cite{hagen2008neutral}, \cite{wickramasuriya2009dynamics}, \cite{gorgas2012quantifying}, \cite{gorgas2012concepts}, \cite{dorninger2013comparison}  & \cite{kumar1993multicomponent}, \cite{briggs1997wavelets}\\
		\midrule[1pt]
		Intensity-Scale Skill Score (ISS) & MRA & recali- brated, dyadic & (categorical) skill score and energy decomposition, skillful scale & \cite{mittermaier2006using}, \cite{mittermaier2008introducing}, \cite{csima2008use}, \cite{shutler2011evaluating}, \cite{hirata2013comparison}, \cite{johnson2014multiscale}, \cite{kumar2013multiscale} & \cite{casati2004wavelet} \cite{casati2007new} \\
		\midrule[1pt]
		ISS revised & MRA & non-dyadic possible & as ISS but with direct bias assessment & \cite{de2011assessing}, \cite{stratman2013use} &  \cite{casati2010new} \\
		\midrule[1pt]
		ISS quantile & MRA & non-dyadic possible & as ISS but based on data-dependent quantile-thresholds & \cite{kwiatkowski2014spatial} & \cite{saux2012wavelet}\\
		\midrule[1pt]
		Wavelet Coefficient Score (WCS) & directional MRA & dyadic & skill score for climate predictions & \cite{dorninger2013comparison} & \cite{livina2008wavelet}\\
		\midrule[1pt]
		Wavelet Spectra Analysis & RDWT & dyadic & Wavelet (Co-) Spectra, scale-dependent weights to minimize total MSE & \cite{bousquet2006analysis} & \cite{turner2004predictability}\\
		\midrule[1pt]
		2D CWT applications to meteorological data & CWT & non-dyadic & denoising, directional scale-separation, feature extraction & & \cite{wang2010two}\\
 		\bottomrule[2pt]\\
\end{tabular}
\caption{A summary of spatial verification methods using wavelet transforms is given including their particular type of wavelet decomposition, requirements on the input data and a list of applications on meteorological data. Possible wavelet transformations are multiresolution analysis (MRA), redundant (or non-decimated) discrete wavelet transform (RDWT) and continuous wavelet transform (CWT).}
\label{tab:method-overview}
\end{table*}

\section{Applications}
\label{sec:applications}

\subsection{PME Techniques}
\label{subsec:pme-applications}

\cite{brunsell2003length} follow the methodology of \cite{kumar1993multicomponent} to study length scales of surface energy fluxes derived from remotely sensed temperature and fractional vegetation data. Wavelet spectra (i.e. the energy of wavelet coefficients for each scale) are calculated from a two-dimensional MRA with the D4 wavelet \citep{daubechies1992ten}. Peaks in the wavelet spectrum, which are associated with dominant scales, are studied.  Temperature and surface fluxes exhibit strong small scale variance, whereas the corresponding wavelet co-spectra do not show noteworthy contributions on small scales. It is argued that this implies a nonlinear relation, and that knowing the dominant scales of remotely sensed data may not be enough to deduce the spatial variability of surface energy fluxes.

\cite{alvera2007forecast} apply EOFs and a MRA similar to \cite{briggs1997wavelets} on spatial fields of sea surface temperature (SST). This allows to identify the scales, which are mainly contributing to the global error. Model output from runs with and without data assimilation are evaluated against observational SST data. The wavelet scale decomposition is able to identify the scales, on which data assimilation leads to the largest error reductions. Furthermore, the localization characteristic of wavelets allows to pinpoint a significant portion of small-scale errors to an area around the Northern Current. It is concluded that a misplacement of the Northern Current strongly contributes to the total error. \cite{smith2011multiscale} follow a similar approach to evaluate atmospheric fields from a three day regional ensemble forecast against analysis data. There, a one-dimensional MRA is applied only along meridians in contrast to the two-dimensional approach by \cite{briggs1997wavelets}. Each scale is evaluated using spread-error plots. Due to the localization capacity of the DWT, areas with the largest error contributions on a particular scale are identified.

\cite{singh2011statistical} use a five-level D6-wavelet \citep{daubechies1992ten} decomposition on geopotential height composite maps. Denoising is carried out analogous to \cite{briggs1997wavelets}. The reconstructed (directional) details on scales 2-5 are then subjected to principal component analysis (PCA) to determine the dominant coefficients. After a training period, these coefficients are used to classify teleconnection patterns associated with significant events at each scale. The statistics of each class are then compared to air pollution data to identify the influence of large-scale weather events on regional air pollution.

\cite{hagen2008neutral} follow the methodology of \cite{briggs1997wavelets} to evaluate the performance of spatial models for landscape change and urban planning with respect to benchmark models. A mother wavelet based on the minimum Shannon entropy criterion is chosen to carry out a two-dimensional MRA to decompose model output and benchmark data sets into orthogonally separated scales. In this context, orthogonality is one of the decisive benefits of the MRA approach, since other common methods of aggregation (e.g. by moving windows \cite{hagen2006map}) allow coarse scale information to remain present on finer scales (i.e. some errors are registered at multiple scales). Sensitivity studies regarding the choice of the mother wavelet as well as the specifics of padding (i.e. adding rows and columns to the image-matrix to obtain quadratic dyadic dimensions) are presented. Although the data used in this study show the island La R\'{e}union (i.e. the padding only adds ocean) a high sensitivity to the positioning of the islands center is found for larger scales. This is a good example for the shift sensitivity of MRA and for issues related to the boundary conditions of discrete sets of data, where DWT usually assumes periodicity. We refer to \cite{mallat1999wavelet} for an elaborate discussion on this topic. \cite{wickramasuriya2009dynamics} apply the methodology of \cite{hagen2008neutral} to a cellular automaton for land-use modeling. The MRA approach exhibits \textit{skillful scales}, i.e. scales where the model at hand outperforms a benchmark model. 

\cite{bousquet2006analysis} use a wavelet decomposition analogous to \cite{turner2004predictability} to verify rainfall accumulations from the HIMAP model of Environment Canada against radar data. First, traditional methods such as MSE, correlation coefficient, PoD, FAR, or ETS are employed. Then, to minimize phase errors, the forecast field is shifted to maximize cross-correlation between forecast and observation. The shifted forecasts are orthogonally decomposed by a two-dimensional Haar-DWT. Wavelet spectra and co-spectra are calculated following \cite{turner2004predictability}. Dominant scales are identified via their spectral peaks. Including the wavelet co-spectra allows to deduce a \textit{maximal predictive scale}. The scale decomposition of the MSE reveals that scales below the maximal predictive scale are responsible for a significant portion of the total error. Following \cite{turner2004predictability}, small scales are filtered out for both forecast and observation by setting their weight $w(l)$ to zero. The resulting fields are then evaluated with the aforementioned traditional method, which is comparable to the methodology of \cite{briggs1997wavelets} with an additional step for shift correction.

\cite{gorgas2012quantifying} study the effect of uncertain observations to verification scores. Spatial fields for wind speed and accumulated precipitation of six NWP models (COSMO-2, COSMO-7, COSMO-DE, COSMO-EU, CMC-GEMH, CMC-GEML) is verified against a multitude of different sets of reference data. The main question is which characteristic of the reference data, e.g. interpolation method, grid resolution or observation density, has the largest impact on verification uncertainty. Traditional verification methods (bias corrected MSE, correlation coefficient) are combined with a two-dimensional Haar-DWT to study verification uncertainty on separate scales. A scale dependency of verification uncertainty is observed, which is larger for low-resolution reference-data grids. A step towards systematic assessment of verification uncertainty is taken in \cite{gorgas2012concepts}, where an analysis ensemble is created employing PCA and wavelet techniques. For the latter, analysis fields are decomposed using two-dimensional DWT with Haar and Coiflet-2 wavelets. The coefficients are filtered with a soft threshold following \cite{briggs1997wavelets}. The remaining non-zero coefficients are then perturbed by Gaussian noise. The inverse 2D-DWT then yields a perturbed analysis field (i.e. one member of the analysis ensemble).

\cite{dorninger2013comparison} evaluate NWP-model chains to assess whether high-resolution models add skill to their driving lower-resolution models and to what extent forecast errors propagate from coarse models to high-resolution models. Multiple different verification methods are employed, e.g. the previously discussed wavelet-based scores \ISS \citep{casati2010new} and WCS \citep{livina2008wavelet} as well as traditional scores (Bias, RMSE, correlation coefficient, standard deviation, centralized Nash-Sutcliffe score) and the object-based method SAL \citep{wernli2008sal}. A two-dimensional Haar-DWT is used to study the traditional scores at separate scales. The hypothesis that high-resolution models outperform their driving low-resolution models could only be partially confirmed. For some verification measures, e.g. the bias-corrected RMSE, the low-resolution models exhibit the best scores. While this may largely be founded in the double penalty issue of pointwise verification methods, scale separation shows that this cannot be the sole reason, since it holds true even on large scales. This study also emphasizes the importance to look at more than one verification measure in order to assess model performance.

\subsection{ISS}
\label{subsec:iss-applications}

\cite{mittermaier2006using} uses the ISS to compare different resolutions of the Unified Model (UM) to assess the benefit of high-resolution model precipitation forecasts. The forecasts are evaluated against radar-rainfall accumulations. The ISS results show that averaging of the raw model output is necessary to achieve positive skill. A modified sign test statistic is introduced to analyze the statistical significance of ISS results. This test is designed to detect the ``presence of persistent error in the intensity-scale phase space'' but could easily be adapted to show significantly skillful areas.

\cite{mittermaier2008introducing} studies the impact of observational uncertainties on forecast skill and error using traditional verification measures such as frequency bias, equitable threat score, ROC curve and log odds ratio, as well as the ISS. 6h and 24h precipitation accumulations of the mesoscale Met Office Unified Model (UM) are evaluated against radar data. Uncertainty is introduced not via the addition of random numbers but by systematically altering the binary masks of radar observations. The unaltered masks are derived with thresholds equal to $2^k$, for $k\in\{-3, \ldots, 7\}$. The threshold values of the observations are then ``lagged'', i.e. equal to $2^{k-1}$, for $k\in\{-3, \ldots, 7\}$. Therefore, the thresholds to derive the binary observation fields are half as large as the ones used for the binary forecast fields. This method has been chosen in order to address potential underestimation of radar observations. Case studies and monthly statistics are analyzed. The introduced threshold lag can be clearly observed in ISS plots, and no unexpected (e.g. scale dependent) effects are present. In general the ISS exhibits a robust behavior to the systematic uncertainties, while some of the traditional scores show more variation. It would be interesting to study the effects of more noisy uncertainties on the ISS.

\cite{csima2008use} apply the ISS to different resolutions of the ECMWF operational QPF over France. Verification is based on gridded precipitation analysis fields obtained from 24h accumulated rain gauge observations. To present the ISS results in a way suitable for operational centers, box-whisker plots for daily ISS versus spatial scale for two different precipitation thresholds are used. No statistically significant difference in ISS values is detected between the two model resolutions.The challenge of very strict requirements of domain characteristics in an operative environment (i.e. dyadic dimension and no missing values) is discussed. This issue is also addressed in \cite{casati2010new}. We would like to note that the verification area in \cite{csima2008use}, i.e. a $16\times 16$ box, is dangerously small for a technique that is based on a translation sensitive MRA, since the boundary will have a significant effect even on small scales. Furthermore, the box-whisker plots demonstrate the necessity to consider statistical significance.

\cite{shutler2011evaluating} combine traditional methods such as RMSE, Frequency bias, Log odds ratio and Kappa coefficient with the ISS and an EOF analysis to evaluate model output for surface concentrations of chlorophyll-a against satellite observations. \cite{hirata2013comparison} follow the approach by \cite{shutler2011evaluating} and compare spatial distributions of phytoplankton functional types on a global scale with satellite observations using bias assessment, correlation statistics, spatial principal component analysis (PCA) and ISS. The wavelet scale separation is able to isolate scales and thresholds of negative skill.

\cite{johnson2014multiscale} examine the impact of perturbations of initial condition and different model physics on precipitation fields obtained with the WRF model. The perturbations are evaluated against a control run, which itself is verified against observations. The methodology follows \cite{casati2004wavelet} with the important difference that real-valued difference fields instead of binary fields are used for the wavelet decomposition. The analysis does not focus on the calculation of skill scores but rather on the decomposition of \textit{perturbation energy}, i.e. the mean square of the precipitation field difference, similar to \cite{casati2007new} and \cite{casati2010new}.

\cite{kumar2013multiscale} follow the ISS methodology to study improvements in surface snow simulations due to topographic adjustments to radiation. Instead of the usual MSE-based skill score, differences in probability of detection (POD) and (negatively oriented) false alarm rates (FAR) are evaluated. During a three year simulation both metrics showed clear improvements for the topographic adjustment. Wavelet scale decomposition pinpoints the majority of these improvements to small scales. 

\cite{de2011assessing} compare seasonal precipitation simulations and inter-annual precipitation differences of the NCEP T62 GCM with results from dynamical downscaling. Both sets of simulations are evaluated against daily precipitation analysis using the \ISS methodology. Energy relative differences are used for an in-depth study of inter-annual precipitation differences. Here, a slightly unconventional perspective is taken: instead of using $E_{rel.diff}$ to compare two models, it is calculated for both models and the observations separately for each region and for two seasons DJF1997 and DJF1998. $E_{rel.diff}$ values from 1997 are subtracted from the ones obtained for 1998 resulting in inter-annual changes in $E_{rel.diff}$, decomposed in spatial scale and intensity. Comparing the results of GCM and RCM with observations yields information about the inter-annual structural change in precipitation events.

\cite{stratman2013use} use \ISS to evaluate reflectivity fields simulated with different initial and boundary conditions of the WRF model. A ``cold start'' run is compared to a ``hot start'' forecast, which assimilates radar data via a 3DVAR/cloud analysis technique. Traditional verification techniques showed a large increase in skill in convective cases. However, based on subjective assessments of many experts, this skill-gain of the hot start model is not justified. It has been suspected that this issue is connected to differences in the frequency bias, which leads to misleading skill scores. \ISS shows that the hot start run exhibits only small improvement on convective scales but large improvements only for spatial scales above 40km, in agreement with the experts' subjective assessment.

\cite{kwiatkowski2014spatial} follow the methodology of \ISSq to assess the ability of earth system models to simulate the spatial patterns of sea surface temperature in important coral reef regions. The models' ability to capture the patterns of monthly SSTs in a historical climatology is studied, as well as the ability to simulate patterns of SST warming anomalies on a climatic timescale.

\section{Texture analysis with wavelets and spatial verification}
\label{sec:texture}
The ISS uses a pointwise MSE on different physical scales to assess forecast skill and error. In cases where we do not expect the spatial forecast and verification fields to correlate in space, we might be, however, interested in assessing the structural similarity of two spatial fields. We thus consider a method for texture analysis \citep{eckley2010locally} based on locally stationary 2D wavelet processes (LS2W). The data is transformed with a Haar-RDWT into a redundant wavelet frame of $3N$ full resolution fields, where $N$ denotes the number of dyadic scales. The main idea is to classify the texture of a field based on its \textit{wavelet spectrum}, i.e.\ the distribution of the energy of wavelet coefficients between scales and directions. \citet{eckley2010locally} show that the \textit{raw wavelet periodogram} (i.e. the spatial mean over squared wavelet coefficients) is a heavily biased estimator for the wavelet spectrum. This is a direct consequence of the redundancy of the RDWT, which leads to non-orthogonal decompositions. An unbiased estimator can be derived by considering the energy leakage between scales and directions via an \textit{inverse local wavelet covariance matrix}; we refer to \cite{eckley2010locally} for mathematical details. This approach reduces the spatial field to a \textit{feature vector} of $3N$ characteristic values. In texture analysis one usually tries to classify (i.e.\ cluster) several sets of textures. In a training period, the texture label for each field is known, so that a \textit{linear discriminant analysis} (LDA), \cite{martinez2001pca,mclachlan2004discriminant}, can be applied to the labeled feature vectors. The LDA projects the $3N$ dimensional data onto linear combination of features, which simultaneously emphasize the differences between two classes and the similarities within each class. The quality of the classification is measured in \textit{singular values} that are defined as the ratio of the outer standard deviation (between clusters) and the inner standard deviation (between members within each cluster).

\cite{marzban2009verification} compare spatial fields with two types of variograms. While the first type is calculated only on nonzero values and evaluates solely the texture of the fields, the second type of variogram is calculated on the whole domain and is sensitive to size and location of objects. Variograms are closely related to autocovariance and autocorrelation, which in turn are closely linked to the power spectrum via Fourier transforms. Since LS2W is based on wavelet periodograms, both methods have much in common but differ in the two important ways: first variograms require some kind of stationarity of the data, e.g. the \textit{intrinsic hypothesis} \citep{matheron1963principles}, while LS2W requires only local stationarity. And second, variograms assume isotropic fields, whereas LS2W is able to resolve directional information. \cite{bosch2004wavelets} provide an in-depth discussion concerning the relationship between wavelets and variograms. Furthermore, generalization of the variogram are considered, which bear the potential to narrow the gap between both approaches.

We applied LS2W to quantitative precipitation forecasts (QPF) of a limited-area, non-hydrostatic numerical weather prediction model. The QPF forecasts particularly aim at the prediction of mesoscale convection. Due to the small spatial scales, mesoscale convection is very difficult to predict, and it is not expected that forecasts provide exact location and timing of a convective event. Verification of QPF on the mesoscale thus aims at an evaluation of the characteristic of the predicted precipitation fields, rather than a verification in terms of MSE or ISS. Texture analysis using wavelet may provide a route towards the verification of structural characteristics of spatial forecasts. In order to demonstrate the potential benefit of texture analysis for spatial verification, we apply this method to an ensemble of QPF, and compare the forecasts to a reanalysis, which are shortly described in the following.

Our forecasts are taken from the COSMO-DE ensemble prediction system (COSMO-DE-EPS) operated by the German Meteorological Service (Deutscher Wetterdienst, DWD). The COSMO-DE-EPS provides 20 forecasts of hourly accumulated precipitation for each hour on a 2.8 km $\times$ 2.8 km grid over Germany and neighboring countries. The forecasts we use are initialized at 0 UTC and predict 21 hours into the future. For a detailed description of COSMO-DE-EPS the reader is referred to \citet{Gebhardt:2011} and \citet{Peralta:2012}, and references therein. 

We compare the QPF to a regional reanalysis (COSMO-REA2) based on a very similar version of the COSMO model. COSMO-REA2 is provided by the Hans-Ertel Centre for Weather Research program\footnote{www.herz-tb4.uni-bonn.de}, and has the same setup as the COSMO-REA6 reanalysis described in \citet{Bollmeyer:2015}.
The only difference between COSMO-REA6 and COSMO-REA2 is the smaller horizontal grid spacing in COSMO-REA2, which does not require a parametrization of deep convection. As in COSMO-REA6, latent heat nudging provides a skillful reanalysis of precipitation. Note that COSMO-REA2 and COSMO-DE-EPS are based on the same limited area model, and thus might share similar systematic biases. However, radar observations have large systematic errors and unobserved areas, whereas gridded analysis based on station observations suffer from interpolation assumptions. Thus, we think that COSMO-REA2 serves the purpose of this application well. The fields are padded with zeros to the next dyadic dimension in such a way that the original field is centered in the larger dyadic field. Note that the issues regarding positional sensitivities \citep{hagen2008neutral,casati2010new} do not arise in this framework, because the RDWT is shift-invariant.

\begin{figure*}
\center
\includegraphics[width=0.8\textwidth]{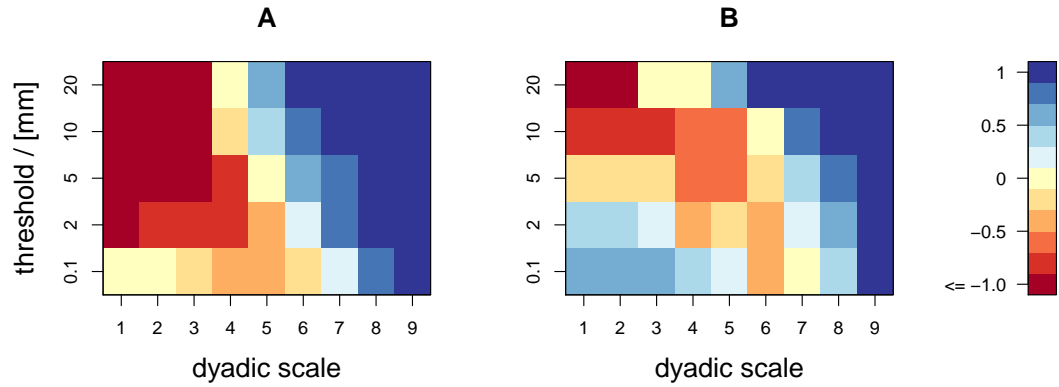}
  \caption{The mean over 4 points in time (14UTC, 16UTC, 18UTC, and 20UTC) and 20 members of the COSMO-DE-EPS of the ISS is plotted as a function of scale and threshold values for 5 June 2011 (A) and 22 June 2011 (B). The regional reanalysis COSMO-REA2 serves as observational data (details can be found in section~\ref{sec:texture}).}
\label{fig:iss}
\end{figure*}

\begin{figure*}
\includegraphics[width=\textwidth]{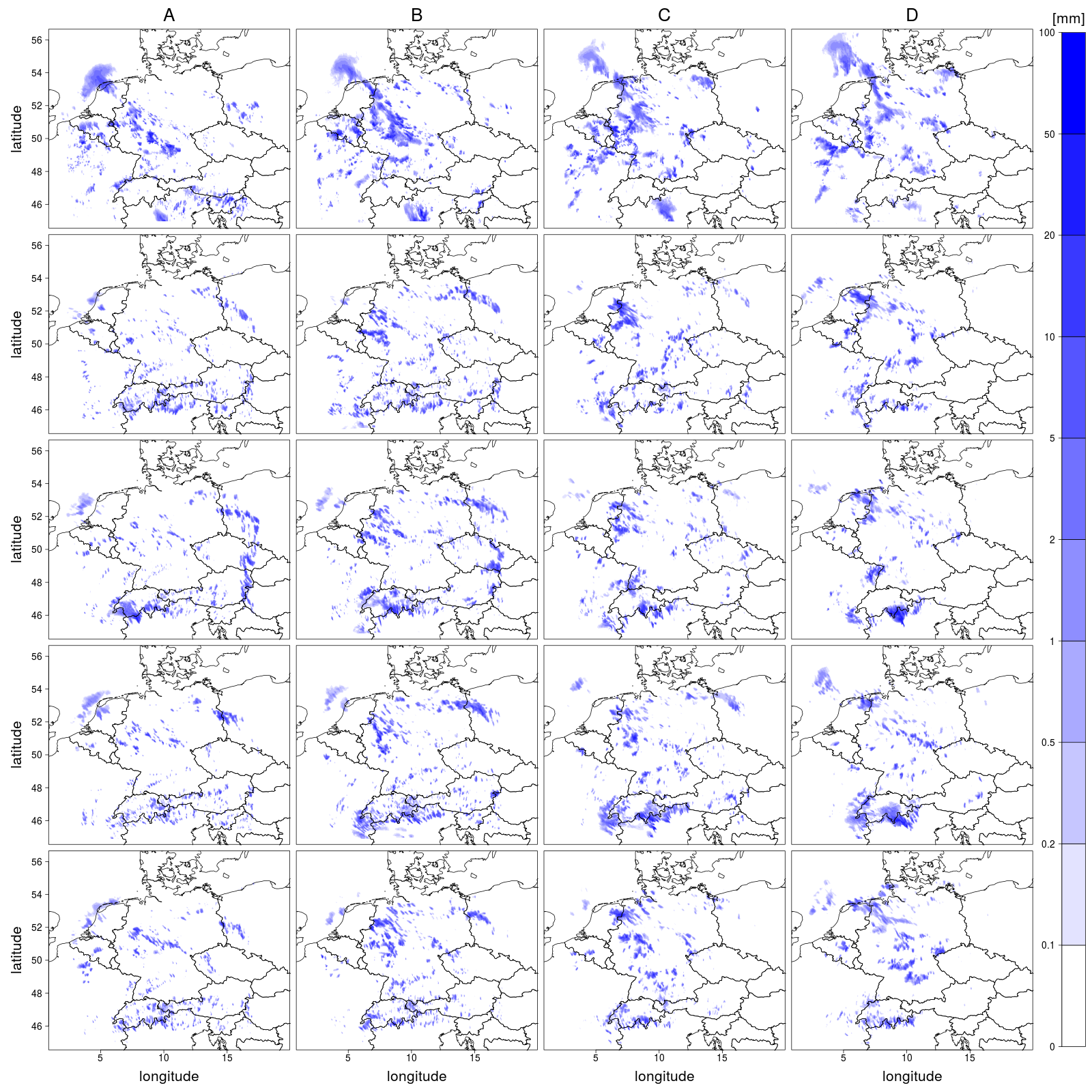}
\caption{Hourly accumulated precipitation in mm per hour for 05 June 2011 at A: 14UTC, B: 16UTC, C: 18UTC, and D: 20UTC (columns) for COSMO-REA2 (upper row) and four selected members (lower four rows).}
\label{fig:precip05june}
\end{figure*}

For our demonstration of the LS2W method, we selected two weather cases, 5 June 2011 and 22 June 2011. During these two days deep moist convection occurred in Germany leading to heavy precipitation and wind gusts. The main difference between the two days is that 5 June 2011 was characterized by scattered and localized convection, whereas during 22 June 2011 a frontal system with deep convection passed over Germany. Details on the synoptic situation and the mesoscale characteristics during the two selected weather cases are discussed in \citet{Weijenborg:2015}.

The reanalysis fields for the hours 14UTC (A), 16UTC (B), 18UTC (C), and 20UTC (D) are displayed in Fig. \ref{fig:precip05june}. Although precipitation shows some structure, the convective cells are scattered over large areas, and not well organized. Four selected ensemble forecasts for each hour are shown in the lower four rows in Fig. \ref{fig:precip05june}. The ensemble forecasts seem to capture the scattering of precipitation, and the regions with precipitation seem to overlap in some places. However, scattering is large and indeed the ISS is only positive on very large scales (Fig. \ref{fig:iss}).

\begin{figure*}
\includegraphics[width=\textwidth]{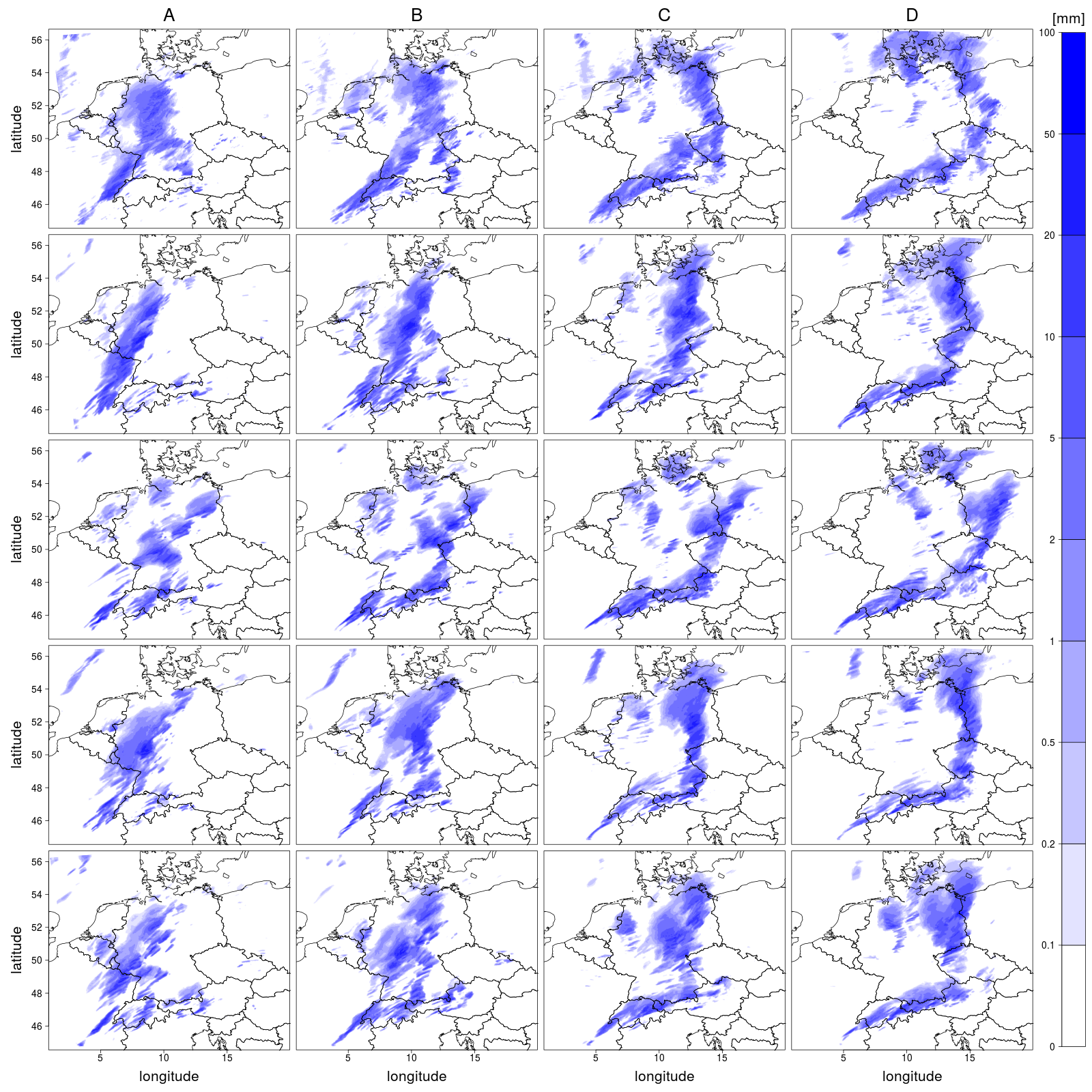}
\caption{Same as Fig. \ref{fig:precip05june} but for 22 June 2015.
}
\label{fig:precip22june}
\end{figure*}

On June 22 2011 the situation is quite different (Fig. \ref{fig:precip05june}). Here, the reanalysis shows a frontal band with heavy precipitation, which is well captured by some of the ensemble predictions. The structure of precipitation is much less scattered. As shown in Fig. \ref{fig:iss}, the ISS is positive for both small and large scales. Fig. \ref{fig:precip05june}  also suggest a preferred direction of propagation from southwest to northeast, which is visible due to the fact that the maps show hourly accumulated precipitation.

We now discuss the application of LS2W for each of the two cases, separately. The first question we ask is whether LS2W is able to detect significant differences with respect to the different points in  time. Since we are interested in structure differences, we did not use the two largest scales, which mainly contain the overall average, for the LDA. It turned out that if included, the largest scales dominate the LD vectors, and separation of the dates is event more significant than without level 9 and 10 shown below. 

\begin{figure*}
\includegraphics[width=0.5\textwidth]{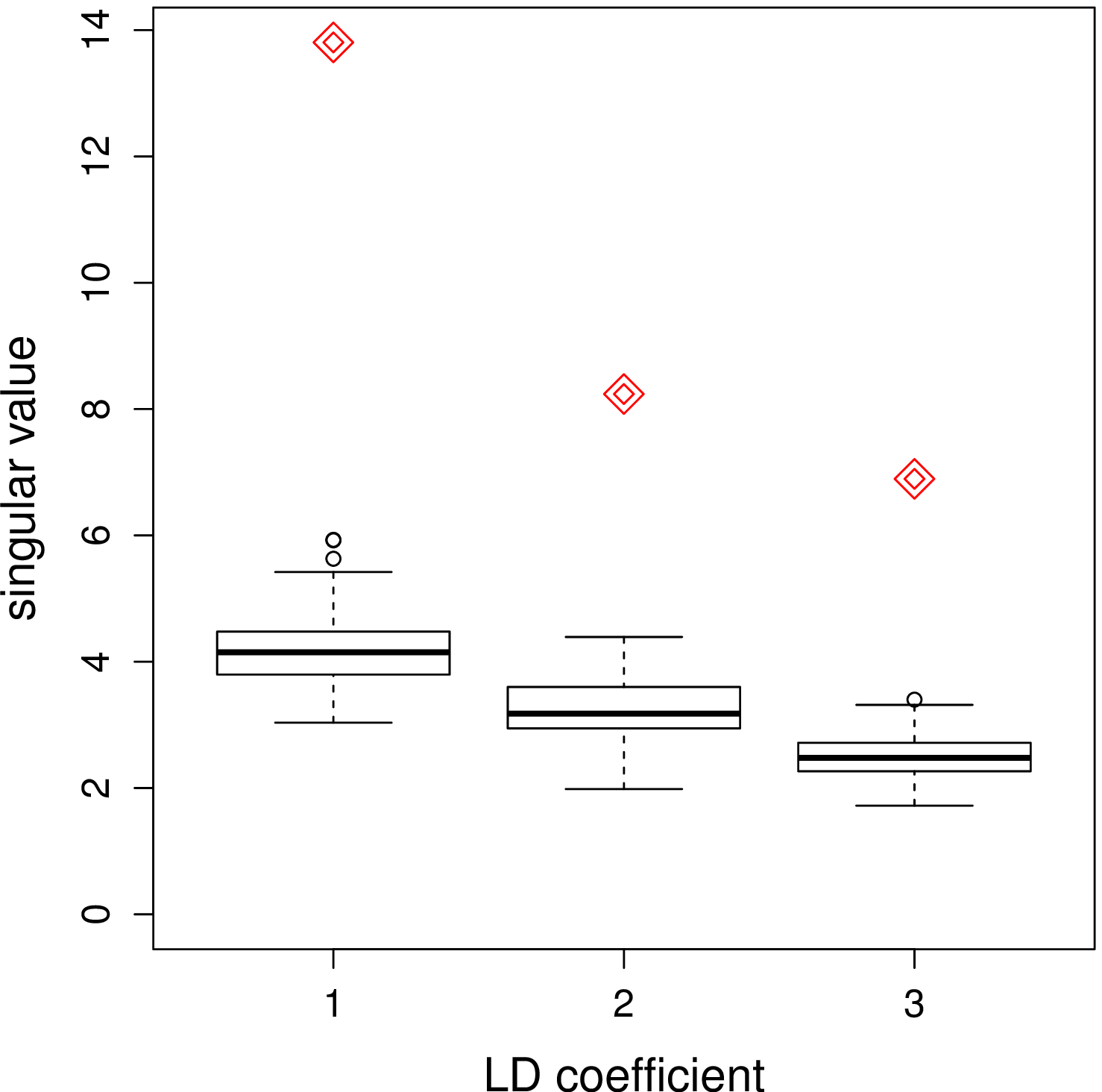}
\includegraphics[width=0.5\textwidth]{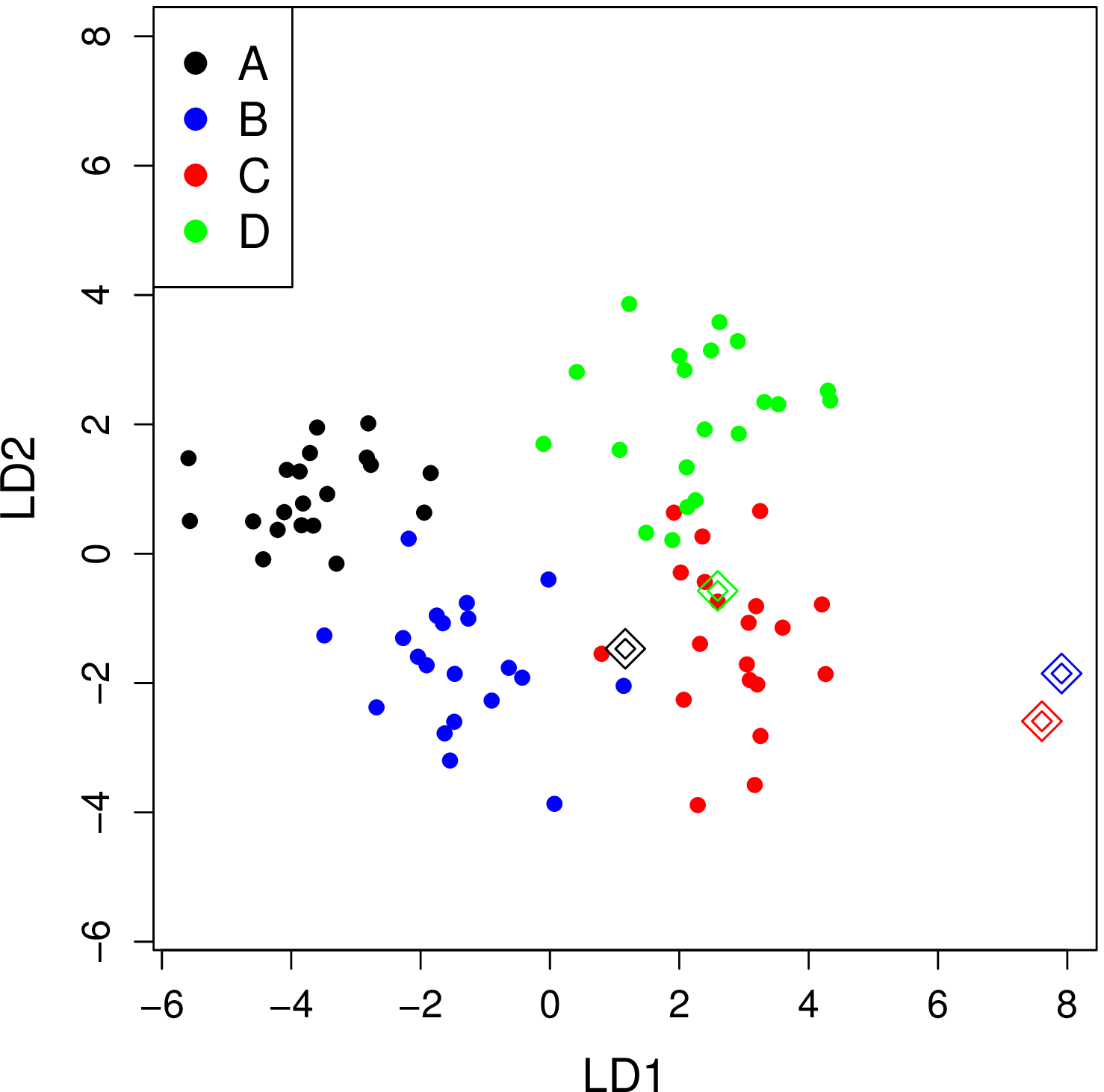}
\caption{Singular values (left) and coefficients of LD vectors (right) from LS2W for 05 June 2015. The singular values are given as red diamond, the box-whisker represent the uncertainty under the null hypothesis of indistinguishable forecasts. The colored dots in the scatter show the first two LD coefficients of the 20 ensemble members for each hour, respectively. The respective coefficients of COSMO-REA2 are given as colored diamonds. The groups represent A: 14UTC, B: 16UTC, C: 18UTC, and D: 20UTC.}
\label{fig:svd_scatter05june}
\end{figure*}

\begin{figure*}
\includegraphics[width=0.5\textwidth]{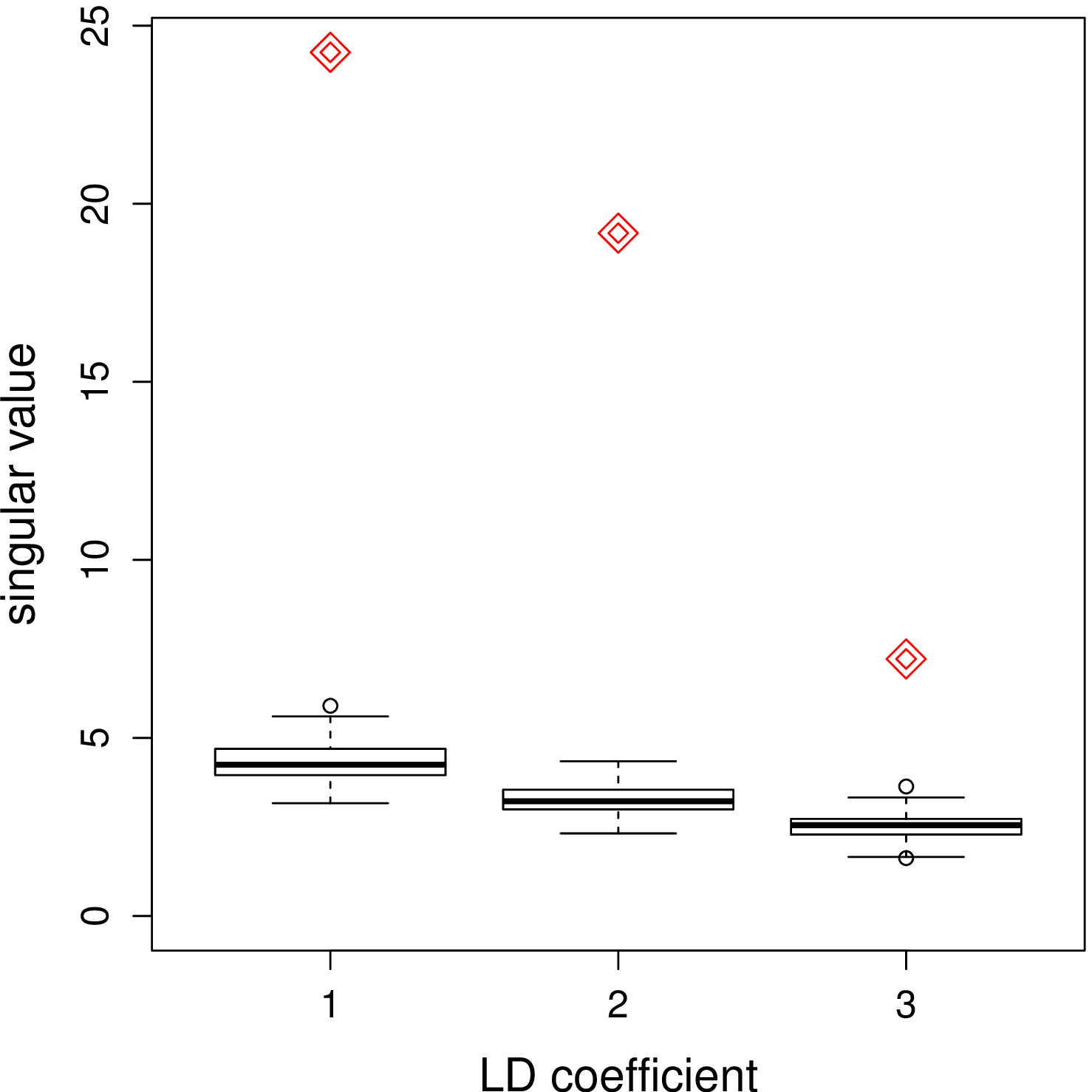}
\includegraphics[width=0.5\textwidth]{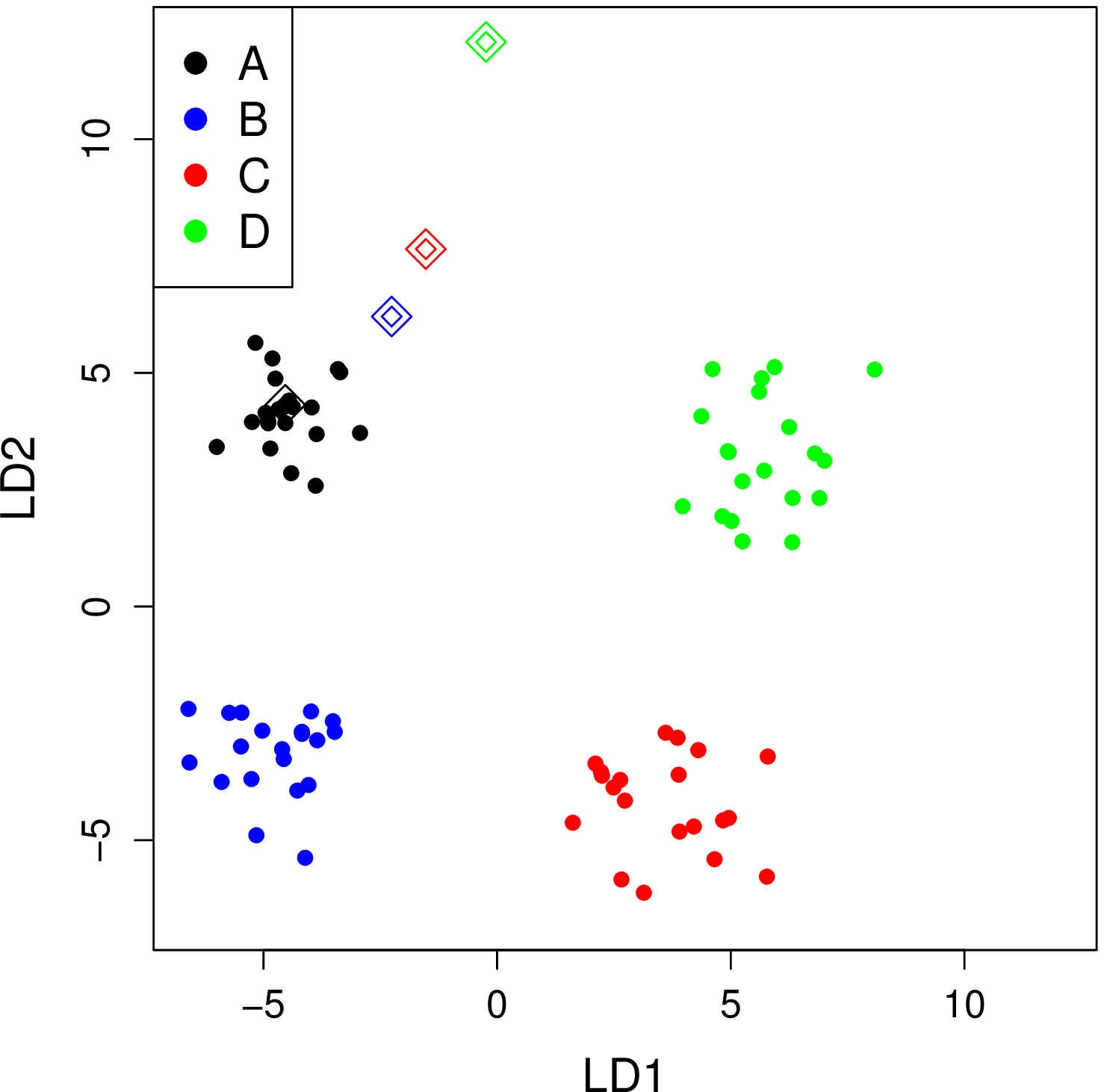}
\caption{Same as Fig. \ref{fig:svd_scatter05june} but for 22 June 2015.}
\label{fig:svd_scatter22june}
\end{figure*}

\begin{figure*}\centering
\includegraphics[width=0.5\textwidth]{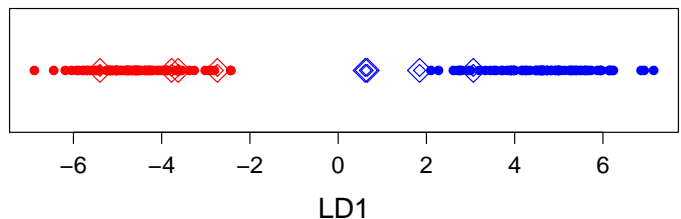}
\caption{Coefficients of LD vectors from LS2W for 05 June 2015 against 22 June 2015. The colored dots show the 4$\times$20  LD coefficients for A: 05 June 2015 and B: 22 June 2015, respectively. The respective coefficients of COSMO-REA2 are given as colored diamonds.}
\label{fig:beide_scatter}
\end{figure*}

\begin{figure*}
\includegraphics[width=0.5\textwidth]{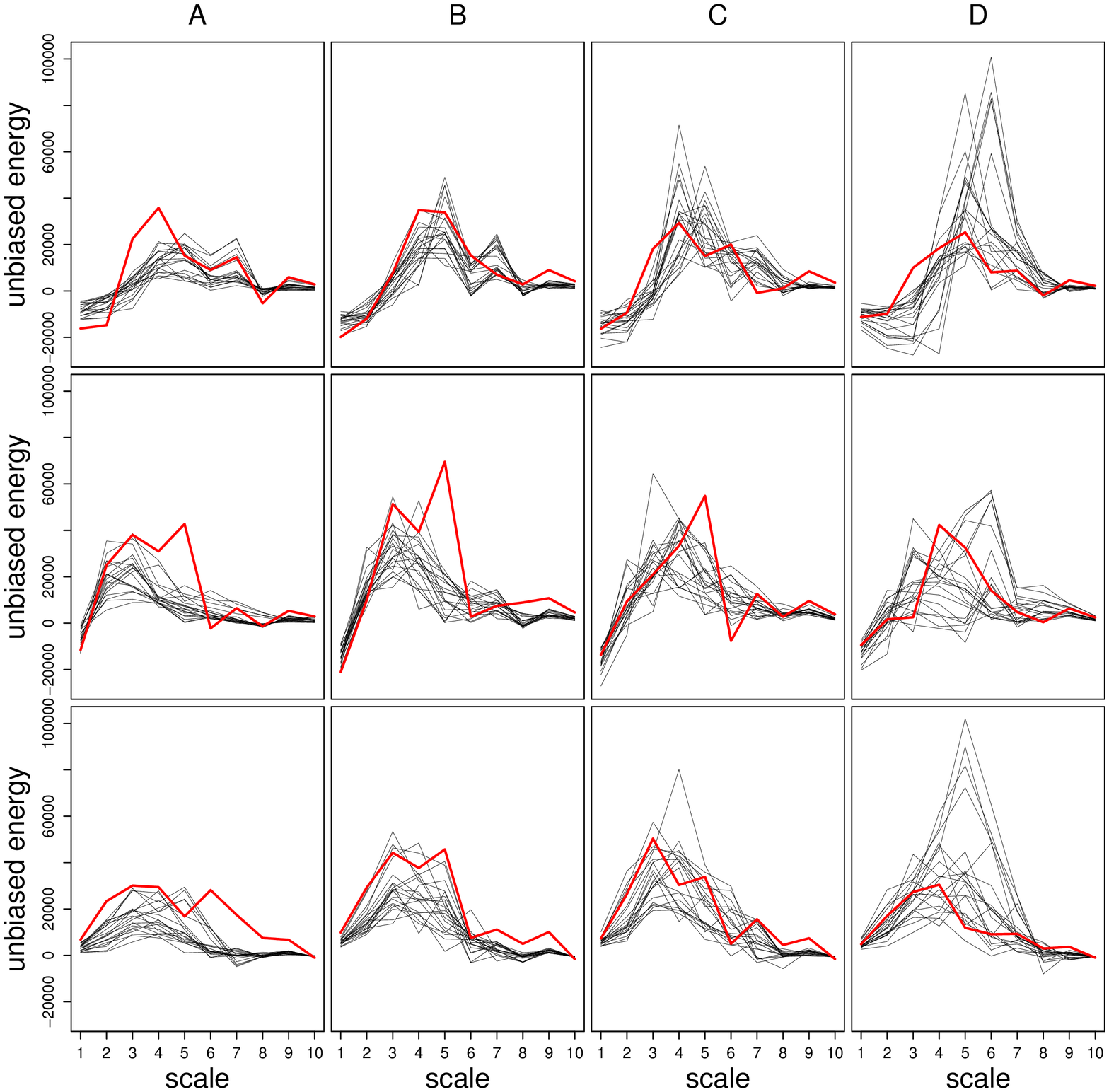}
\includegraphics[width=0.5\textwidth]{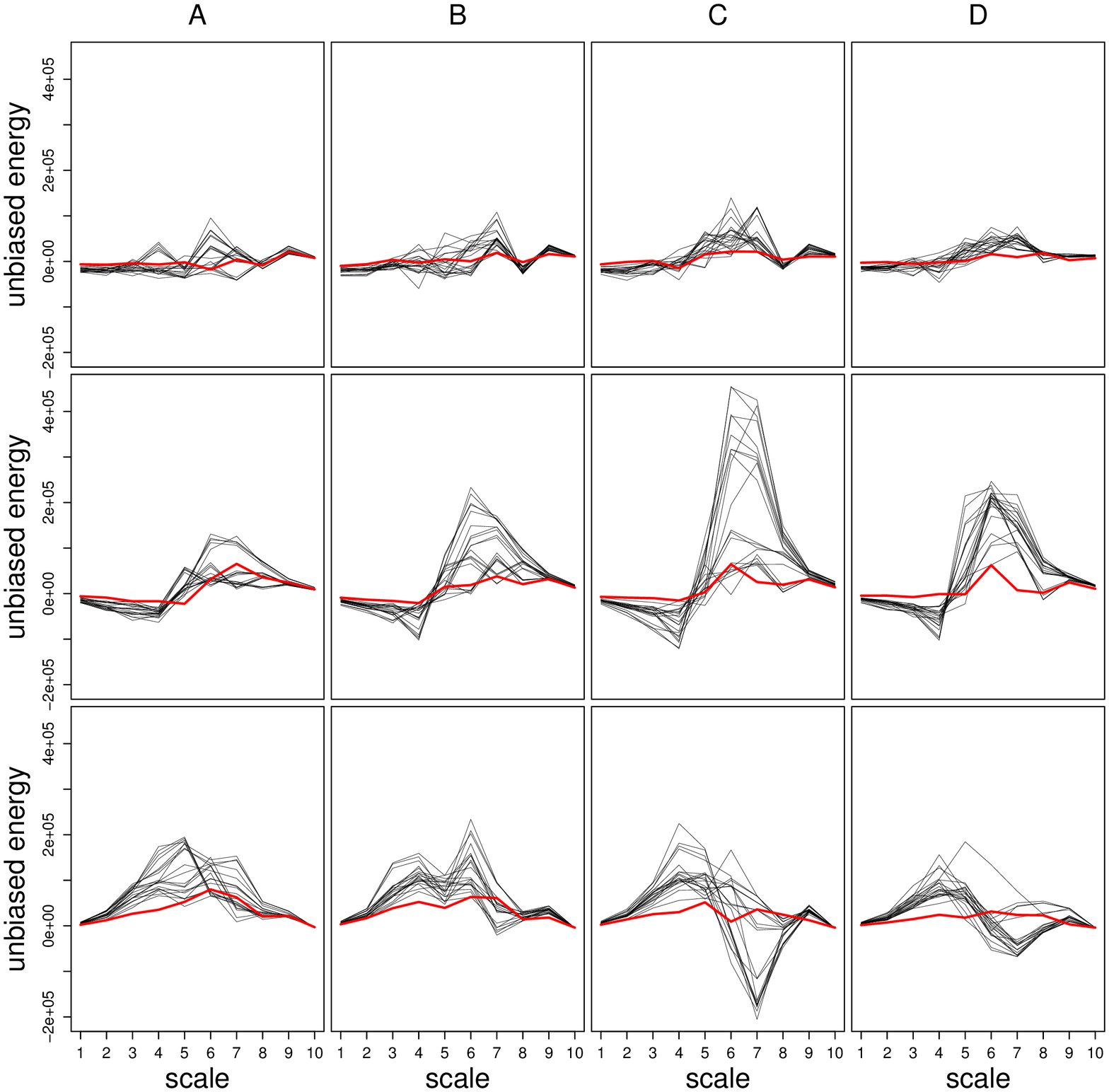}
\caption{Unbiased estimates of the 2D wavelet spectra for 05 June 2015 (left) and 22 June 2015 (right). The rows represent the vertical (upper row), horizontal (middle), and diagonal (lower row) direction. The black lines represent the energy of the 20 members of COSMO-DE-EPS, and the red line COSMO-REA2.  reanalysis. The columns represent A: 14UTC, B: 16UTC, C: 18UTC, and D: 20UTC.}
\label{fig:perio_beide}
\end{figure*}

We start with 05 June 2011. Fig. \ref{fig:svd_scatter05june} shows the singular values of the three LD vectors. Note that the LD vectors are derived using the ensemble forecasts only, thus each group A to D contains 20 precipitation forecasts for one hour, respectively. The box-whiskers represent a bootstrap uncertainty for randomly chosen groups (i.e.\ the hours are perturbed randomly). It seems that LS2W is able to significantly separate the different hours of the forecast, since the SVD largely exceed the uncertainty under the nullhypothesis of indistinguishable forecasts.

Fig. \ref{fig:svd_scatter05june} shows the LD coefficients of the leading 2 LD vectors. The figure confirms that the forecasts for the different forecast hours are distinguishable. This is even more clear if the third LD vector is considered (not shown). 
We then use the 3 LD vectors to predict the group (i.e.\ hour of observation) for the reanalysis. Unfortunately, all COSMO-REA2 fields are predicted to represent group C, also seen in \ref{fig:svd_scatter05june}, where two of the four LD coefficients of COSMO-REA2 lie outside the range spanned by the QPF fields. 

For 22 June 2011 the LDA is able to distinguish different forecast hours even better, which can be seen in the scatter plot of Fig. \ref{fig:svd_scatter22june}. The SVD values are also larger than for 05 June 2011 and support the visual impression. However, a classification of the reanalysis with respect to the forecast hour is not successful, since all reanalysis fields are classified to belong to A. 

Although a significant separation of the ensemble QPF of different hours seems possible, a correct classification of the respective observations is not. We thus performed the same analysis with only two groups A and B, where A contains all 4$\times$20 QPF of 05 June 2011, and B all 4$\times$20 QPF of 22 June 2011. Fig. \ref{fig:beide_scatter} shows the LD coefficient of the LD vector. It is not surprising that the two days are well separable. But, also the COSMO-REA2 are now well classified, and seem to be represented by the COSMO-DE-EPS forecasts, at least in the  one-dimensional space spanned by the LD vector.

The unbiased periodograms which form the basis of the feature vectors in LS2W are displayed in Fig. \ref{fig:perio_beide} for both days and each hour, respectively. Largest difference in the periodograms can be detected in the diagonal components, where on 05 June 2011 most energy is present on the smaller scales (levels 2-4), whereas on 22 June 2011, energy is largest on levels 6-8 (large scales). However, the interpretation of the LD vectors (not shown) and the unbiased periodograms needs further research, as well as the design of appropriate and meaningful scores.

\section{Conclusions}
\label{sec:conclusions}
Wavelet transforms offer an effective framework to decompose spatial data into separate (and possibly orthogonal) scales. During the last decade methods such as point measure enhancements (PME), the Intensity-Scale Skill Score (ISS) and the Wavelet Coefficient Score (WCS) were developed based on two-dimensional multiresolution analysis (MRA) decompositions, and the number of meteorological applications, particularly of the ISS, has seen significant growth in the last five years. Most of the existing methods assess forecast skill or error on each scale using tradition pointwise verification measures. Their focus lies on the evaluation of different resolutions of the original data in image-space. Results from other scientific fields such as feature detection, image fusion, texture analysis, or facial and biometric recognition show that more sophisticated wavelet transforms such as \textit{Redundant Discrete Wavelet Transforms} (RDWT), \textit{Dual-Tree Complex Wavelet Transform} (DTCWT) and Continuous Wavelet Transforms (CWT) contain considerable potential to derive useful diagnostic information. This is particularly true if the interpretation of results is not restricted to the image space (low-pass information), but includes the analysis of wavelet coefficients (high-pass information).
Although wavelet techniques are usually referred to as scale separation techniques they have much in common with neighborhood techniques, when their focus lies on low-pass filters (e.g. PME, ISS). This is particularly true for the Haar wavelet, since its scaling function is equivalent to simple box averaging, which leads to verification scores that are easy to interpret. For more sophisticated wavelet families the low-pass filter can be understood as directionally weighted averaging. On the other hand, the analysis of wavelet coefficients (i.e. focusing on high-pass filters) is closely connected to aspects of feature based methods, such as texture analysis or key point detection. Therefore, wavelet based spatial verification techniques can potentially combine diagnostic information from different methodological branches.

We have applied a technique developed for texture analysis (LS2W) in the context of high-resolution quantitative precipitation forecasting. Using \textit{Linear Discrepant Analysis} (LDA) on unbiased wavelet periodograms allowed for an analysis and statistically robust clustering of structural characteristics of quantitative precipitation forecast (QPF) ensemble data.

\ack We gratefully acknowledge financial funding by the project High Definition Clouds and Precipitation for advancing Climate Prediction HD(CP)$^2$ funded by the German Ministry for Education and Research (BMBF) under grant FK 01LK1209B. 


 \bibliographystyle{wileyqj}
 \bibliography{weniger_bib}

\end{document}